\documentclass[reprint,superscriptaddress,showpacs,nofootinbib,amsmath,amssymb,aps,longbibliography,prb]{revtex4-2}
\usepackage{graphicx}
\usepackage{dcolumn}
\usepackage{bm}
\usepackage[utf8]{inputenc}
\usepackage[english]{babel}
\usepackage{braket}

\preprint{APS/123-QED}

\begin{document}

\title{Heisenberg-Langevin theory of an exciton mirror}

\author{S. V. Andreev}
\email[Electronic adress: ]{Serguey.Andreev@gmail.com}
\affiliation{Mediterranean Institute of Fundamental Physics}

\date{\today}

\begin{abstract}

We develop a Heisenberg-Langevin theory of an exciton mirror accounting for the retardation and the long-range electron-hole exchange. The particular case of a single-component exciton gas is analyzed in detail. The optical bistability due to repulsion between the excitons inside the light cone appears to be prone to a modulational instability towards the non-radiative polariton modes. Above the corresponding threshold, the pumped 2D exciton gas acts as an optical parametric generator of twin polariton beams. Conversely, below the threshold, the mirror acquires the phase-conjugating properties.                                           

\end{abstract}


\maketitle

\section{Introduction}

In the realm of non-linear optics, mirrors have traditionally served as indispensable yet passive components of devices  \cite{Boyd}. The miniaturization and the pursuit for strong photon correlations at ultra-low powers has challenged this paradigm. With the advent of semiconductor technology, an appealing concept known as the exciton mirror has emerged \cite{Ivchenko2004}, wherein resonant reflectance and non-linearity may be combined at the ultimate scale of a single atomic layer \cite{Wild2018, Back2018, Scuri2018}. An exciton is a hybrid of a photon and electron-hole excitations \cite{Agranovitch}, and the Coulomb forces between the fermions may enable pair correlations between the photons down to the limit of vanishing density. Recent experiments (Ref. \cite{Scuri2018} and references therein) have provided compelling evidence of the third-order [$\chi^{(3)}$] exciton non-linearity in a monolayer: a significant change of the reflectivity (or transmission) upon variation of the exciton density, reminiscent of the optical bistability in a Fabry-Perot cavity with a Kerr medium \cite{Gibbs}.

On the theoretical side, the Heisenberg–Langevin approach \cite{Scully} has become a fundamental tool for describing nonlinear phenomena in mesoscopic quantum systems. In the input–output formalism of Gardiner and Collett \cite{Gardiner1985}, this approach is supplemented by boundary conditions that relate incoming and outgoing photon-field operators, thereby providing a second-quantized scattering description of open photonic systems. Within this approach, the $\chi^{(3)}$ optical bistability may be readily obtained at the mean-field level in a steady state \cite{Zeytinoglu2017}.

The non-linear behaviour of the exciton mirror reported in Ref. \cite{Scuri2018}, however, pertains to a pulsed regime. To date, no signatures of steady-state dispersive optical bistability have been reported in this or related \cite{Back2018} work. Moreover, even in the pulsed regime, the observed dependence of the reflectance on the input power lacks the characteristic hysteresis loop \cite{Scuri2018}. The latter feature would be essential for implementing exciton mirrors as optical memory cells \cite{Gibbs}.

In this work, we develop a specific input-output theory accounting for the retardation and the long-range electron-hole exchange within an exciton \cite{OpticalOrientation, Pekar1958, Bir1970, Nakayama1985, Cho1999, Andreani1990, Sham, Gupalov, Glazov, Qiu2015, Nestoklon2018, Becker2018, Andreev2021, Andreev2022_1}, producing dramatic changes of the single-exciton dispersion. In a broader context, the fine structure of the exciton levels due to the long-range exchange is also known as the longitudinal-transverse (TE-TM) splitting \cite{Agranovitch}. Our equations constitute a basis for understanding of collective phenomena in driven 2D excitons and provide an insight into the mirror bistability issue. We perform an instructive analysis of a circularly-polarized exciton gas, where the two-body interactions are repulsive. We find that the hitherto assumed picture of an exciton condensate depleted by interactions \cite{Zeytinoglu2017} is restricted to a narrow radiative region of in-plane momenta ("light cone"), and the bistability appears to be prone to a stimulated decay towards the non-radiative polariton modes. The counter-propagating polariton beams emerge in a steady state above certain threshold, may possess extremely high group velocities (approaching $c$) and may produce output signals with perfect correlation in their intensities - the so-called "twins" \cite{Klyshko1967, Heidmann1987}. Hysteresis and high reflectivity at normal incidence, on the other hand, would require tailoring the polariton decay channels. Our theory allows one to deduce a corresponding upper bound on the polariton lifetime along the dispersion curve. We also demonstrate that, below the instability threshold, the exciton mirror acquires the phase-conjugating properties \cite{Zeldovich1985}, and may be used to amplify and rectify the 2D signals (\textit{e.g.}, guided polaritons or plasmons) spoiled by disorder.

\section{Input-output formalism for a 2D excitons gas}

We begin with a generic quantum-electrodynamic description of 2D excitons. Consider a transition-metal-dichalcogenide (TMD) monolayer\footnote{Our consideration of a single atomic layer may be extended to other planar structures, such as epitaxial \cite{Ivchenko2004, Andreani1990, Sham, Gupalov} or colloidal \cite{Nestoklon2018, Becker2018} quantum wells.} occupying a large square quantization area $S$. The polarization density associated with the exciton spin component $\sigma$ is written as
\begin{equation}
\label{Polarization}
\hat{\bm P}_\sigma(\bm r,z)=\frac{1}{\sqrt{S}}\sum_{\bm k}\psi_{1s}^\ast(0)\bm d_\sigma\hat b_{\sigma,\bm k}e^{i\bm k\bm r}\lvert\chi(z)\rvert^2+h.c.,
\end{equation}
where $\bm r=(x,y)$ is the in-plane coordinate and $\chi(z)$ describes confinement of the constituent fermions in the out-of-plane direction. For simplicity, we assume one and the same envelope both for the electron and the hole transverse motion. The function $\chi(z)$ will enter the theory through the Fourier transform
\begin{equation}
\label{OverlapIntegral}
i_{cv}(q_z)\equiv \int \lvert\chi(z)\rvert^2e^{iq_z z}dz,
\end{equation}
also known as the overlap integral \cite{Ivchenko2004}. The envelope $\psi_{1s}(\rho)$ of the electron-hole $s$-wave relative motion in the structure plane is taken at $\rho=0$ \cite{Elliott1957}. The excitons possess massive spatial dispersion $\omega_{\sigma}(\bm k)=\omega_{\sigma}(0)+\hbar k^2/2m$ inherited from the fermions.

The exciton spin may be described by the spin-$1$ operator of a transverse electromagnetic field \cite{Andreev2022_1}
\begin{equation}
\label{BosonSpin}
\hat{\bm s}=\hat\sigma_x \bm n_x+\hat\sigma_y\bm n_y+\hat \sigma_z \bm n_z,
\end{equation}
where $\hat\sigma_x$, $\hat\sigma_y$ and $\hat\sigma_z$ are Pauli matrices. We have adopted the notation $\ket{\uparrow}$ and $\ket{\downarrow}$ for the basis states characterized by $s_z=+1$ and $s_z=-1$, respectively. These states correspond to two orthogonal polarizations of the in-plane electric dipole
\begin{equation}
\label{ElectricDipole}
\bm d_{\uparrow,\downarrow}=\mp\frac{\bm e_x\pm i\bm e_y}{\sqrt{2}}\frac{ep_{vc}}{i\omega_{\uparrow,\downarrow}(\bm k)m},
\end{equation}          
with the $\pm$ signs standing for $\sigma=\uparrow,\downarrow$ and $p_{vc}$ being the inter-band momentum matrix element. The bosonic excitations described by the operators $\hat b_{\uparrow,\bm k}$ and $\hat b_{\downarrow,\bm k}$ are occasionally referred to as "bright" excitons, and they govern radiative properties of a vast majority of semiconductor nanostructures \cite{Andreani1990, Sham, Gupalov, Ivchenko2004, Nestoklon2018, Becker2018}. For the TMD monolayer, the spin label $\sigma$ may be identified with the valley index \cite{Xiao2012, Glazov, Andreev2022_1}. Notably, under time reversal, $t\rightarrow -t$, one has
\begin{equation}
\label{TimeReversal}
\hat s_{x,y}\rightarrow \hat s_{x,y},
\end{equation}
as expected for the linear polarization \cite{Andreev2022_1}. 

By contrast, since $\hat s_{z}\rightarrow -\hat s_{z}$, one expects the exciton doublet to exhibit a Zeeman splitting, $\omega_{\downarrow}(\bm k)-\omega_{\uparrow}(\bm k)=2\mu_B \mathsf{g} B_z/\hbar$, in an external magnetic field $\bm B=B_x \bm n_x+B_y\bm n_y+B_z \bm n_z$. Such a splitting has indeed been observed in 2D semiconductors \cite{Zeng2012, Sallen2012, Srivastava2015, Stier2016}. We shall make use of this effect below, where it will motivate our consideration of a single-component exciton gas.

The monolayer is embedded in a quantization box of volume $V=SL$, $L\rightarrow\infty$, hosting the photon field
\begin{equation}
\hat{\bm E}(\bm r, z)=i\sum_{\bm k,q_z,\alpha}\sqrt{\frac{\hbar\omega_{\bm q}}{2V\epsilon_0}}\hat a_{\alpha,\bm q}\bm e_{\bm q, \alpha} e^{i\bm (\bm k\bm r+q_z z)}+h.c.
\end{equation}
which is decomposed into 3D plane waves with wave vectors $\bm q=(\bm k, q_z)$ and pairs of mutually orthogonal transverse polarizations labelled by $\alpha$. The second-quantized electric-dipole Hamiltonian of the system reads                     
\begin{widetext}
\begin{equation}
\label{PZW}
\begin{split}
\hat H&= \sum_{\bm q,\alpha}\hbar\omega_{\bm q}\hat a_{\bm q\alpha}^\dagger\hat a_{\bm q\alpha}+\sum_{\bm k,\sigma}\hbar\omega_\sigma(\bm k)\hat b_{\sigma\bm k}^\dagger\hat b_{\sigma\bm k}+\frac{1}{2S}\sum_{\bm k_1,\bm k_2,\bm q,\sigma,\sigma^\prime}\hat b_{\sigma^\prime, \bm k_2-\bm q}^\dagger \hat b_{\sigma, \bm k_1+\bm q}^\dagger g_{\sigma\sigma^\prime} \hat b_{\sigma,\bm k_1}\hat b_{\sigma^\prime,\bm k_2}\\
&-\sum_\sigma\int  \hat{\bm E}(\bm r,z)\cdot \hat{\bm P}_\sigma(\bm r,z)d\bm r dz+\frac{1}{4\epsilon_0}\sum_{\sigma,\sigma^\prime} \int  \hat{\bm P}_{\sigma^\prime,\perp}^\dagger  \hat{\bm P}_{\sigma,\perp} d\bm r dz,
\end{split}
\end{equation}
\end{widetext}
where the last term is the dipole self-energy. The mechanical-exciton part includes two-body interactions. By virtue of the optical selection rules, a pair of bosons with $s_{z,1}=s_{z,2}=\pm1$ (total spin $S_z=\pm2$) exhibits triplet spin configurations for the constituent pairs of identical fermions, so that the ensuing boson-boson interaction is predominantly repulsive. At low collisional energies, such interactions are fully characterized by the $s$-wave scattering lengths $a_{\sigma\sigma}$ being on the order of the corresponding microscopic force ranges. In the dilute limit $na_{\sigma\sigma}^2\ll 1$, such that $\left |\ln(na_{\sigma\sigma}^2)\right |\gtrsim 1$ \cite{Wang2026} (or, ultimately, $\left |\ln(na_{\sigma\sigma}^2)\right |\gg 1$), the effective two-body interaction potentials may be estimated as $g_{\sigma\sigma}=-4\pi\hbar^2/m \mathrm{Re}[(\ln(na_{\sigma\sigma}^2)-i\pi)^{-1}]$ \cite{Utesov2024, Andreev2024}. Consistently, we shall assume $g_{\sigma\sigma}>0$. The same expression may be used for $g_{\uparrow\downarrow}$, although for a weakly-bound biexciton one may have $na_{\uparrow\downarrow}^2\gg 1$. In reality, the biexcitons are tightly bound due to exchange within singlets of the identical fermions, so that, again, $na_{\uparrow\downarrow}^2\ll 1$, and one has $g_{\uparrow\downarrow}>0$ (see further remarks in Appendix \ref{sec:parameters}).

\begin{figure}
\includegraphics[width=1\linewidth]{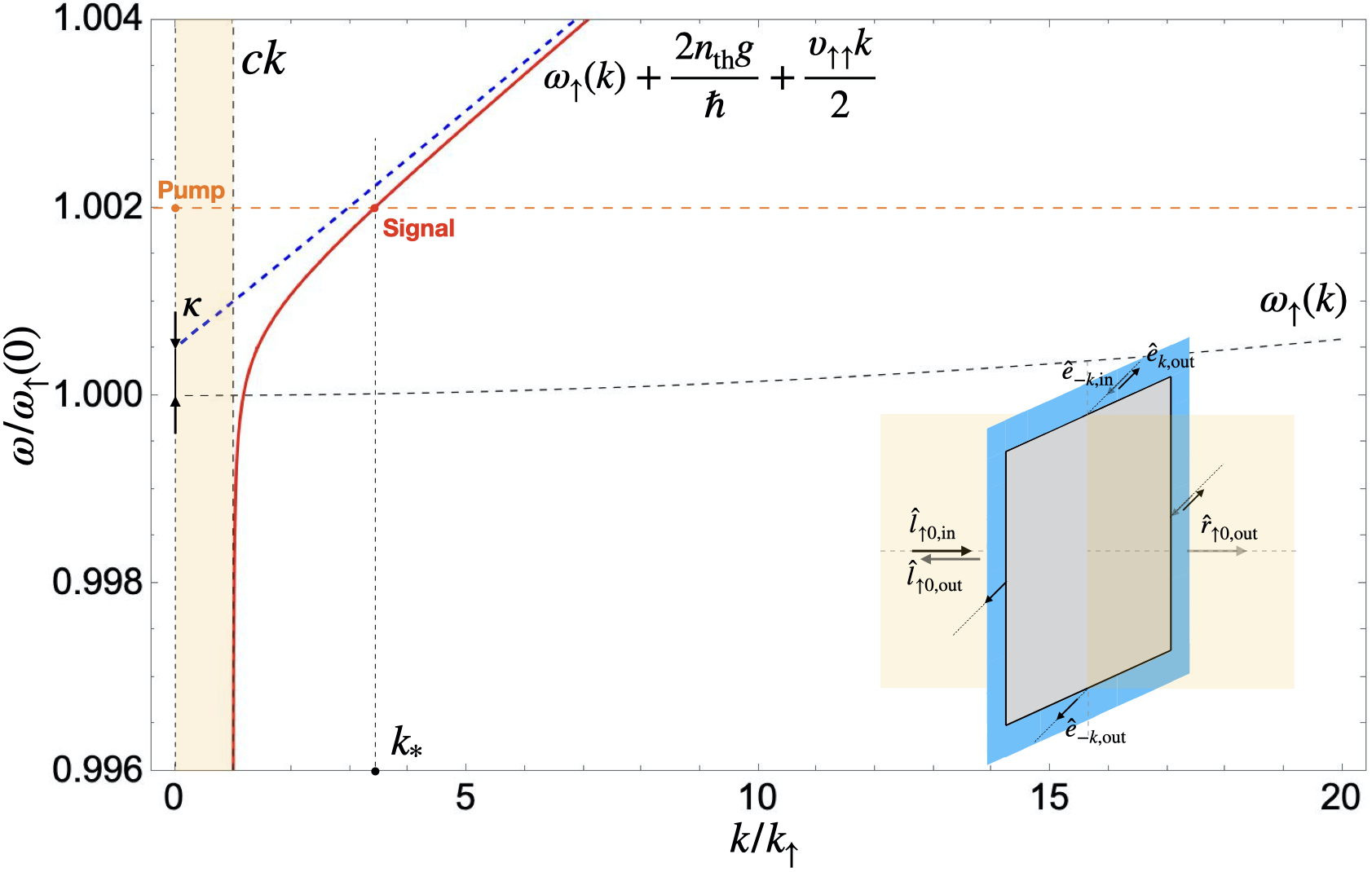} 
\caption{\label{Fig1} The polariton dispersion $\varepsilon_\uparrow(\bm k)$ (red line) at the instability threshold. The parameters typical for the MoSe$_2$ monolayer have been used. Grey dashed lines show the mechanical exciton dispersion $\omega_\uparrow(\bm k)$ and the boundary of the light cone $ck$. Blue dashed line depicts the large-$k$ asymptote of $\varepsilon_\uparrow(\bm k)$. The wave vector of the instability is defined by the resonance condition $\varepsilon_\uparrow(\bm k_\ast)=\omega$. The inset shows configuration of the incident, reflected and transmitted pump ($\hat l_{\uparrow 0, \mathrm{in}}$, $\hat l_{\uparrow 0, \mathrm{out}}$ and $\hat r_{\uparrow 0, \mathrm{out}}$) and in-plane edge ($\hat e_{-\bm k, \mathrm{in}}$, $\hat e_{\bm k, \mathrm{out}}$ and $\hat e_{-\bm k, \mathrm{out}}$, respectively) fields.}
\end{figure}

By proceeding within the rotating wave approximation along the lines of Ref. \cite{Gardiner1985}, one may derive the set of coupled Heisenberg-Langevin equations of motion for the exciton operators,
\begin{equation}
\label{ExcHeisenbergLangevinEq}
\begin{split}
\frac{d\hat b_{\sigma\bm k}}{dt}&=-i\sum_{\sigma^\prime}[\omega_\sigma(\bm k)\delta_{\sigma\sigma^\prime}-\Delta_{\sigma\sigma^\prime}]\hat b_{\sigma^\prime\bm k}\\
&-\frac{i}{\hbar}\sum_{\sigma^\prime,\bm q}\frac{g_{\sigma\sigma^\prime}}{S}\sum_{\bm p}\hat b_{\sigma^\prime,\bm p+\bm q}^\dagger\hat b_{\sigma^\prime\bm p}\hat b_{\sigma,\bm k+\bm q}\\
&-\sum_\alpha \left[\sqrt{\gamma_{\sigma\alpha}^{(+)}(\bm k)}\hat l_{\bm k\alpha,\mathrm{in}}(t)+\sqrt{\gamma_{\sigma\alpha}^{(-)}(\bm k)}\hat r_{\bm k\alpha,\mathrm{in}}(t)\right]\\
&-\frac{1}{2}\sum_{\alpha,\sigma^\prime,\nu=\pm}\sqrt{\gamma_{\sigma^\prime\alpha}^{(\nu)\ast}(\bm k)\gamma_{\sigma\alpha}^{(\nu)}(\bm k)}\hat b_{\sigma^\prime \bm k}(t),
\end{split}
\end{equation}
supplemented with the continuity relations
\begin{equation}
\label{BoundaryCond1}
\begin{split}
&\sum_\alpha \sqrt{\gamma_{\sigma\alpha}^{(-)}(\bm k)}\left[\hat l_{\bm k\alpha,\mathrm{out}}(t)-\hat r_{\bm k\alpha,\mathrm{in}}(t)\right]\\
&+\sum_\alpha\sqrt{\gamma_{\sigma\alpha}^{(+)}(\bm k)}\left[\hat r_{\bm k\alpha,\mathrm{out}}(t)-\hat l_{\bm k\alpha,\mathrm{in}}(t)\right]\\
&=\sum_{\alpha,\sigma^\prime,\nu=\pm}\sqrt{\gamma_{\sigma^\prime\alpha}^{(\nu)\ast}(\bm k)\gamma_{\sigma\alpha}^{(\nu)}(\bm k)}\hat b_{\sigma^\prime \bm k}(t)
\end{split}
\end{equation}
and the unitarity constraints
\begin{equation}
\label{BoundaryCond2}
\begin{split}
&\sqrt{\gamma_{\sigma\alpha}^{(-)}(\bm k)}\left[\hat l_{\bm k\alpha,\mathrm{out}}(t)-\hat r_{\bm k\alpha,\mathrm{in}}(t)\right]\\
&=\sqrt{\gamma_{\sigma\alpha}^{(+)}(\bm k)}\left[\hat r_{\bm k\alpha,\mathrm{out}}(t)-\hat l_{\bm k\alpha,\mathrm{in}}(t)\right].
\end{split}
\end{equation}
The input (output) photon fields obey the commutation rules
\begin{equation*}
\begin{split}
[\hat l_{\alpha\bm k,\mathrm{in}(\mathrm{out})}(t),\hat l_{\alpha^\prime\bm k^\prime,\mathrm{in}(\mathrm{out})}^\dagger(t^\prime)]&=\delta_{\alpha\alpha^\prime}\delta_{\bm k\bm k^\prime}\delta (t-t^\prime)\\
[\hat r_{\alpha\bm k,\mathrm{in}(\mathrm{out})}(t),\hat r_{\alpha^\prime\bm k^\prime,\mathrm{in}(\mathrm{out})}^\dagger(t^\prime)]&=\delta_{\alpha\alpha^\prime}\delta_{\bm k\bm k^\prime}\delta (t-t^\prime)
\end{split}
\end{equation*}
and describe the in- or out-going analytic signals in the left or right half-spaces evaluated at the plane hosting the excitons. Their intuitive representation is provided in Fig. \ref{Fig1}.

\IfFileExists{Fig1.jpeg}
  {\typeout{=== 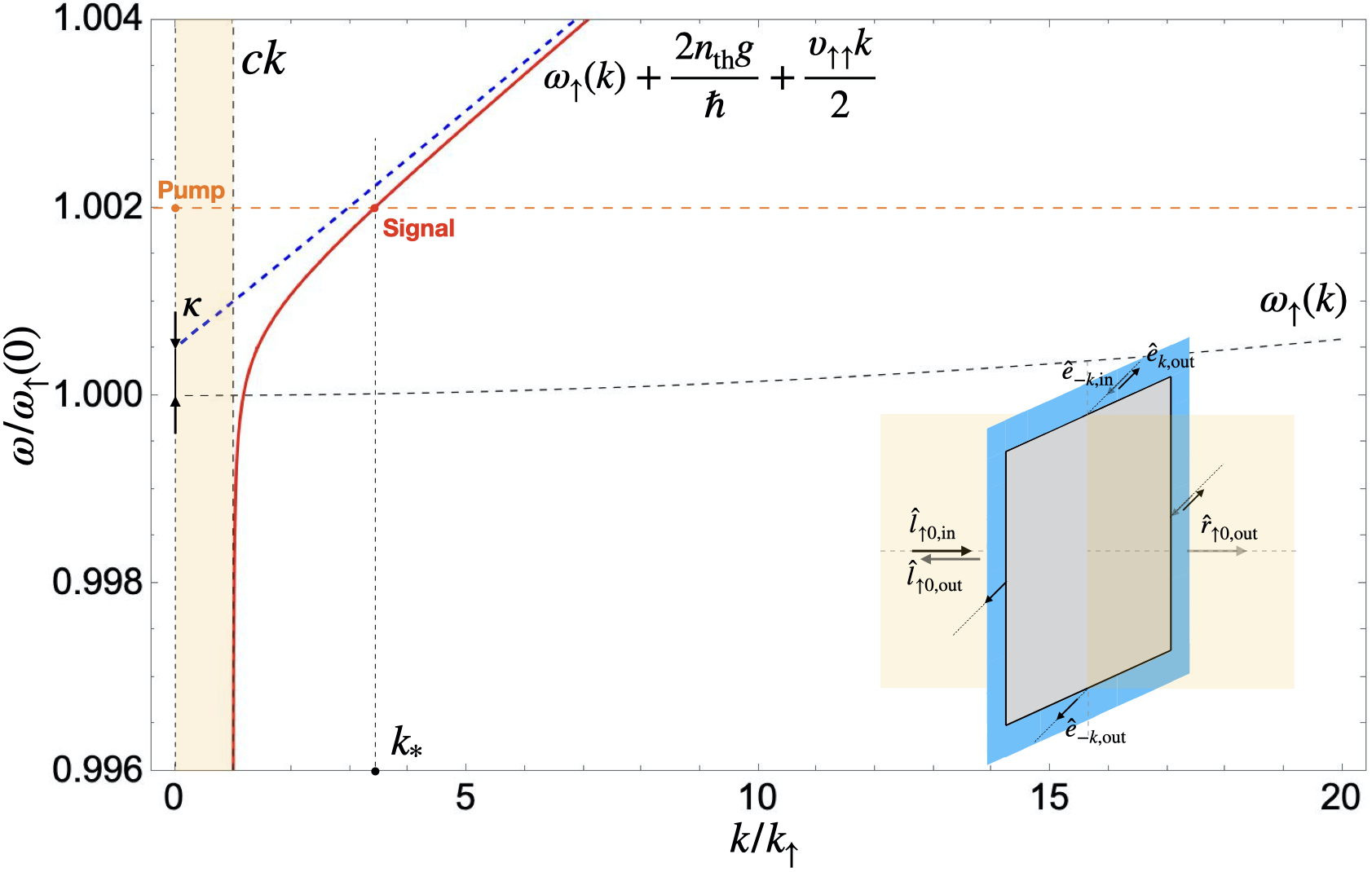 FOUND ===}}
  {\typeout{=== Fig1.jpeg NOT FOUND ===}}

The couplings 
\begin{equation}
\label{GenericRadiativeRate}
\sqrt{\gamma_{\sigma\alpha}^{(\pm)}(\bm k)}\equiv \sqrt{2\pi D_{\bm k}(\varepsilon)}g_{\alpha\sigma}^{(\pm)}(\varepsilon)
\end{equation}
define the radiative decay rates in the last term of Eq. \eqref{ExcHeisenbergLangevinEq}, which are related to the frequency shifts (Lamb shifts) by the Kramers-Kronig relation 
\begin{equation}
\label{LambShift}
\Delta_{\sigma\sigma^\prime}(\varepsilon)\equiv\mathrm{V.p.}\int d\omega\frac{\varepsilon D_{\bm k}(\omega)}{\omega(\omega-\varepsilon)}\sum_{\alpha,\nu=\pm}g_{\alpha\sigma^\prime}^{(\nu)\ast}(\omega)g_{\alpha\sigma}^{(\nu)}(\omega),
\end{equation}
with 
\begin{equation}
\label{PhotonDOS}
D_{\bm k}(\omega)=\frac{L}{2\pi}\int\limits_{0}^{\infty} dq_z\delta(\omega-\omega_{\bm q})
\end{equation}
being the (half-space) photon density of states (DOS) and
\begin{equation}
\hbar g_{\alpha\sigma}^{(\pm)}(\omega)\equiv \sqrt{\frac{\hbar\omega}{2L\varepsilon_0}}\psi_{1s}(0)i_{cv}(q_\omega)\bm e_{\bm k,\alpha}^{(\pm)}\cdot\bm d_\sigma^\ast.
\end{equation}
Here $q_\omega\equiv\sqrt{\omega^2/c^2-k^2}$, the overlap integral $i_{cv}(q_z)$ is defined by Eq. \eqref{OverlapIntegral}, and the unit polarization vector $\bm e_{\bm k,\alpha}^{(\pm)}$ is the short-hand notation for $\bm e_{\bm q\alpha}$ of a monochromatic input field with $\bm q=(\bm k, \pm q_\omega)$. 

The modified exciton dispersion $\varepsilon(\bm k)$ is a self-consistent solution of the corresponding stationary eigenvalue problem. It comprises two branches, which we continue to label by $\sigma$ for convenience, although the corresponding states may, in general, differ substantially from the circular-polarization basis. As a first approximation, one may put $\varepsilon_\sigma(\bm k)\equiv \omega_\sigma(\bm k)$ in Eq. \eqref{GenericRadiativeRate}. For $k>k_\sigma\equiv \omega_\sigma(k_\sigma)/c$, the photon DOS becomes identically zero, and there is no radiative decay. The coupling of excitons to the input photon fields $\hat l_{\bm k\alpha,\mathrm{in}}(t)$ and $\hat r_{\bm k\alpha,\mathrm{in}}(t)$ also disappears. The Heisenberg-Langevin equations \eqref{ExcHeisenbergLangevinEq} become hermitian. The proper expressions for the energy shifts $\Delta_{\sigma\sigma^\prime}$ in this case may be obtained by substituting Eq. \eqref{PhotonDOS} into Eq. \eqref{LambShift} and exchanging the order of integration. One may then remove the principal value prescription ($\mathrm{V.p.}$). It is possible to get a simple analytical result by putting $i_{cv}(q_z)\equiv 1$, which corresponds to the Dirac $\delta$-function approximation for the envelope $\chi(z)$. One obtains
\begin{equation}
\label{EnergyShifts}
\begin{split}
\Delta_{\uparrow\uparrow}(\varepsilon)&=\frac{\upsilon_{\uparrow\uparrow}}{2c}\left (\frac{{\varepsilon}^2}{\sqrt{k^2c^2-\varepsilon^2}}-\sqrt{k^2c^2-\varepsilon^2}\right)\\
\Delta_{\uparrow\downarrow}(\varepsilon)&=\frac{e^{-2 i\theta}\upsilon_{\uparrow\downarrow} k^2}{2\sqrt{k^2-\varepsilon^2/c^2}},
\end{split}
\end{equation}
where $\theta$ is the angle between the in-plane exciton momentum $\bm k$ and the $x$-axis. We have defined $\upsilon_{\sigma\sigma^\prime}\equiv c\gamma_\sigma(\omega_\sigma)/\omega_{\sigma^\prime}$, where the quantities
\begin{equation}
\label{RadiativeRate}
\gamma_\sigma(\omega)=\frac{\omega}{\omega_\sigma}\frac{e^2|p_{cv}|^2|\psi_{1s}(0)|^2}{2c\varepsilon_0\hbar\omega_\sigma m^2}
\end{equation}
would yield the radiative decay rates at normal incidence for $\omega=\varepsilon$ and $\bm k=0$ \cite{Hanamura1988}. At $k\gg k_\sigma$ the retardation effect in Eqs. \eqref{EnergyShifts} may be neglected, and the \textit{single-particle} part of the Heisenberg-Langevin equations \eqref{ExcHeisenbergLangevinEq} takes the useful form \cite{OpticalOrientation, Bir1970, Glazov, Andreev2021, Andreev2022_1}
\begin{equation}
\label{PolaritonEq}
\frac{d\hat b_{\sigma\bm k}}{dt}=-i\sum_{\sigma^\prime}\left[\left(\omega_\sigma(\bm k)+\frac{\upsilon_\sigma k}{2}\right)\delta_{\sigma\sigma^\prime}-\bm\Omega (\bm k)\cdot \bm s_{\sigma\sigma^\prime}\right]\hat b_{\sigma^\prime\bm k},
\end{equation}
where the off-diagonal term has the form of interaction of the boson spin \eqref{BosonSpin} with the 2D effective magnetic field (described by its Larmour frequency)
\begin{equation*}
\bm\Omega (\bm k)=-\frac{\upsilon_{\uparrow\downarrow}k}{2}(\bm n_x \cos 2\theta+\bm n_y \sin 2\theta).
\end{equation*}
The field $\bm\Omega (\bm k)$ is an even function of the exciton momentum $\bm k$, consistently with the time-reversal symmetry of $\hat{\bm s}$ [Eq. \eqref{TimeReversal}].

\section{Bistability and instability of an exciton mirror}

\subsection{Single-component excitons gas} 

The Zeeman splitting of the exciton doublet in an ordinary magnetic field, on the other hand, provides physical motivation for an analysis of the limiting case $\mu_B \mathsf{g} |B_z|\gg \hbar\upsilon_{\uparrow\downarrow}k$ with $k$ being the relevant momentum scale, where the coupling between the two spin components [as exemplified by $\bm\Omega (\bm k)$ above] can be disregarded. Despite such drastic simplification,\footnote{Noteworthy, while elimination of an exciton spin component may be warranted, it is still crucial to keep two-component structure of the electromagnetic field. An exciton $\hat b_{\sigma\bm k}$ with finite $\bm k$ couples to both photon polarizations, as one may readily see, \textit{e.g.}, from Eq. \eqref{LambShift}. At $k\gg k_\sigma$ virtual coupling of this kind produces linear exciton dispersion [see Eq. \eqref{PolaritonEq} and Fig. \ref{Fig1}], which has recently been observed experimentally \cite{Liu2026}.} an analysis of a single-component gas we present below is non-trivial, and it allows one to develop a feeling for both bistability and instability phenomena of the exciton mirror. We shall return to the full two-component problem in the end of the paper.

By introducing the simplified notations, $\hat b_{\bm k}\equiv \hat b_{\uparrow, \bm k}$, $\hat l_{\bm k, \mathrm{in}}\equiv\hat l_{\bm k \uparrow, \mathrm{in}}$ and $g_{\uparrow\uparrow}\equiv g>0$,\footnote{Note, however, that in a sufficiently strong magnetic field along $z$, the interaction constant $g_{\uparrow\uparrow}$ for the lowest energy spin-polarized excitons may be tuned from positive to negative values by applying an in-plane uniaxial strain and sweeping the magnetic field across the biexciton Feshbach resonance \cite{Andreev2024}.} we obtain from Eq. \eqref{ExcHeisenbergLangevinEq}
\begin{equation}
\label{HLequation}
\begin{split}
\frac{d \hat b_{\bm k}}{dt}&=-i[\omega_\uparrow(\bm k)-\Delta_{\uparrow\uparrow}(\bm k)] \hat b_{\bm k}-\sqrt{\gamma_{\uparrow\uparrow}^{(+)}(\bm k)}\hat l_{\bm k, \mathrm{in}}\\
&-\sqrt{\gamma_{\uparrow\uparrow}^{(-)}(\bm k)}\hat r_{\bm k, \mathrm{in}}-\gamma_{\bm k}\hat b_{\bm k}-\frac{ig}{\hbar S}\sum_{\bm q,\bm p}\hat b_{\bm p+\bm q}^\dagger\hat b_{\bm p}\hat b_{\bm k+\bm q},
\end{split}
\end{equation}
where, for $k<k_\uparrow=\omega_\uparrow(k_\uparrow)/c$, we have defined
\begin{equation}
\label{gamma_k}
\gamma_{\bm k}\equiv\frac{1}{2}\gamma_\uparrow(\omega)\frac{q}{q_\omega}\left (1+\frac{q_\omega^2}{q^2}\right)\Bigr |_{\omega=\omega_\uparrow}
\end{equation}
with $\gamma_\uparrow(\omega)$ given by Eq. \eqref{RadiativeRate}. Assuming monochromatic excitation at normal incidence, $\hat l_{\bm k, \mathrm{in}}(t)\equiv \delta_{\bm k, 0} \hat{\mathsf l}_\mathrm{in}e^{-i\omega t}$, one obtains a steady-state mean-field solution
\begin{equation}
\label{CondensateSteady}
\mathsf b_0=-\frac{\sqrt{\gamma_0}}{i[\omega_\uparrow(0)-\Delta_{\uparrow\uparrow}(0)+gn_0/\hbar-\omega]+\gamma_0}\mathsf l_\mathrm{in}
\end{equation}
with $n_{\bm q}\equiv\tfrac{1}{S}\braket{\hat b_{\bm q}^\dagger\hat b_{\bm q}}$.
The input-output relations \eqref{BoundaryCond1} and \eqref{BoundaryCond2} in this particular case take the simple form
\begin{subequations}
\begin{align}
\label{InputOutput}
\mathsf l_\mathrm{out}&=\sqrt{\gamma_0}\mathsf b_0\\
t-r&=1,
\end{align}
\end{subequations}
with $r= l_\mathrm{out}/l_\mathrm{in}$ and $t= r_\mathrm{out}/l_\mathrm{in}$ being the amplitude reflexion and transmission coefficients, respectively. By combining Eq. \eqref{CondensateSteady} with Eq. \eqref{InputOutput} and introducing the detunings $\bar\Delta_0\equiv \omega_\uparrow(0)-\Delta_{\uparrow\uparrow}(0)-\omega$ and $\Delta_0\equiv \bar\Delta_0+gn_0/\hbar$, one may derive the following equation for $\Delta_0$:
\begin{equation}
\label{Bistability}
|r(\Delta_0)|^2=\frac{\hbar \gamma_0(\Delta_0-\bar\Delta_0)}{g |\mathsf l_\mathrm{in}|^2}.
\end{equation}       
The very same equation may be found in description of the optical bistability phenomenon in a Fabry-Perot cavity with a Kerr medium \cite{Gibbs}. The possible solutions for $\Delta_0$, and, hence, the exciton density $n_0$ correspond to intersections of the straight line and the bell-shape contour of $|r(\Delta_0)|^2$. The variation of these solutions with the input intensity $|\mathsf l_\mathrm{in}|^2$ yields the characteristic hysteresis curve for $|r|^2$. Remarkably, the optically generated exciton gas possesses both resonant feedback and a Kerr-like nonlinearity, so that a single atomically thin mirror can exhibit optical bistability without forming a conventional cavity.

The hysteresis may be shown to require $|\bar\Delta_0|>\sqrt{3}\gamma_0$. There is an optimum detuning $\bar \Delta_\mathrm{min}=-2\gamma_0$ yielding the minimum density 
\begin{equation}
\label{LowerDensity}
n_\mathrm{min}^{(-)}=\frac{\hbar\gamma_0}{g}
\end{equation}
for the switching from low to high reflectivity. The $100$ $\%$ reflectivity can be achieved by moving in the opposite direction of the hysteresis curve (\textit{i. e.}, by lowering $|\mathsf l_\mathrm{in}|^2$) at $n_\mathrm{min}^{(+)}=5\hbar\gamma_0/3g$.

The steady-state solution \eqref{CondensateSteady} is stable against conversion into finite-momentum exciton modes with $k<k_\uparrow$, for which Eq. \eqref{gamma_k} gives $\gamma_{\bm k}\geqslant \gamma_0$. For $k>k_\uparrow$, however, this inequality no longer holds. In a disorder-free monolayer, the corresponding polariton modes would have infinite lifetimes, leading to an instability at arbitrarily small condensate density $n_0$. To allow a finite range of $n_0$, where the bistable regime identified above could be observed, we now go beyond the model considered so far and introduce a finite decay rate for the non-radiative modes. In practice, outcoupling of these modes into the far field can be engineered using, for example, a grating \cite{Tsesses2018}, prism coupling \cite{Hao2025}, or coupling to plasmons \cite{Zhou2017,Jin2026}. On the theoretical side, we keep the treatment as simple as possible by embedding our optically pumped sample into a larger (infinite) monolayer medium, which is kept in the dark. The boundaries of the sample are assumed to be imperfect; one may imagine nanometre-scale gaps \cite{Nikitin2013}, which give rise to partial transmission \cite{Maradudin1999, Nikitin2007, Rejaei2015}. The limit of full transmission (no gap) corresponds simply to an excitation spot of size $S$ on an extended monolayer. This construction plays a dual role: the imperfect boundaries broaden the polariton modes, while the surrounding medium supports freely propagating fields that will serve as polariton input and output signals, represented by $\hat e_{\bm k, \mathrm{in}}$ and $\hat e_{\bm k, \mathrm{out}}$, respectively. Thus, for $k>k_\uparrow$, we replace the terms $[\sqrt{\gamma_{\uparrow\uparrow}^{(+)}(\bm k)}\hat l_{\bm k, \mathrm{in}}+\sqrt{\gamma_{\uparrow\uparrow}^{(-)}(\bm k)}\hat r_{\bm k, \mathrm{in}}]$, $\gamma_{\bm k}\hat b_{\bm k}$ by $\sqrt{\kappa_{\bm k}}\hat e_{\bm k, \mathrm{in}}$, $\tfrac{\kappa_{\bm k}}{2}\hat b_{\bm k}$, respectively, in Eq. \eqref{HLequation}. One may then show, that implementation of the optical bistability at $\bm k=0$ would require
\begin{equation}
\label{BistabilityCriterion}
\begin{split}
\frac{\kappa_{\bm k}}{2}\geqslant \gamma_0-\frac{\varepsilon_{\uparrow}(\bm k)-\varepsilon_{\uparrow}(0)}{\sqrt{3}},& \:\:\ \sqrt{3}\gamma_0\geqslant\varepsilon_{\uparrow}(\bm k)-\varepsilon_{\uparrow}(0)\geqslant 0\\
\frac{\kappa_{\bm k}}{2}\geqslant \gamma_0,&\:\:\ \varepsilon_\uparrow(\bm k)< \varepsilon_{\uparrow}(0).
\end{split}
\end{equation}                               

We shall now consider the case, where this requirement is violated. For simplicity, we assume constant $\kappa_{\bm k}=\kappa$ such that $\kappa<2\gamma_0$. We expect, in particular, that $\kappa\ll 2\gamma_0$ is the most natural situation realized in suspended monolayers. Let us fix the detuning $\bar\Delta_0$ and gradually increase the density $n_0$ by increasing the excitation intensity $|\mathsf l_\mathrm{in}|^2$. At the threshold density
\begin{equation}
\label{ThresholdDensity}
n_\mathrm{th}=\frac{\hbar \kappa}{2g}
\end{equation}
the input power starts to convert into the sidebands at $\pm \bm k_\ast$. The magnitude of the wave vector $\bm k_\ast$ is defined by the resonance condition
\begin{equation}
\label{OptimalDetuning}
\varepsilon_{\uparrow}(\bm k_\ast)=\omega_\uparrow(\bm k_\ast)-\Delta_{\uparrow\uparrow}(\bm k_\ast)+\frac{2n_0 g}{\hbar}=\omega
\end{equation}
with $n_0=n_\mathrm{th}$, as shown schematically in Fig. \ref{Fig1}. As an example, we have chosen the pump frequency $\omega$ such that $\bar\Delta_0=\bar\Delta_\mathrm{min}=-2\gamma_0$, the detuning that minimizes the lower turning-point density of the bistable response [see Eq. \eqref{LowerDensity}]. Even at this optimal detuning, however, the finite-momentum instability sets in before the bistable regime can be reached: Eq. \eqref{ThresholdDensity}, together with our assumption $\kappa<2\gamma_0$, implies $n_\mathrm{th}<n_\mathrm{min}^{(-)}$.       

In the first approximation, the density $n_0$ of the $\bm k=0$ mean-field background saturates at $n_\mathrm{th}$. The sum of the phases of $\mathsf b_{\pm\bm k}$ is locked to $\mathsf b_0$,
\begin{equation}
\label{PhaseRelation}
2\phi_0-\phi_{\bm k}-\phi_{-\bm k}=\frac{\pi}{2}\:\:\:\:\ (\mathrm{mod} \:2\pi),
\end{equation}
as one can readily deduce from the steady-state condition $\frac{d}{dt}|\mathsf b_{\pm\bm k}|^2=0$. The polariton wave amplitude grows as $| b_{\pm\bm k}|\sim\sqrt{|\mathsf l_\mathrm{in}|-|\mathsf l_\mathrm{in}^{(\mathrm{th})}|}$, suggesting the phenomenology of a second-order (quantum) phase transition. The relative phase $\phi_{\bm k}-\phi_{-\bm k}$ would be chosen spontaneously and would define the position of a density wave (stripe) with respect to the origin. Depending on the actual geometry, various types of lattices are possible. Such spatially ordered states could be observed in the near field \cite{Tsesses2018}. On the other hand, in the limit of full transmission each of the constituent fields, $\hat b_{\bm k}$ and $\hat b_{-\bm k}$, respectively, may be detected separately due to their leakage through the corresponding edges.\footnote{In this case the operators $\hat b_{\pm\bm k}$ should, strictly speaking, be understood as the polariton envelopes averaged over the illuminated region.} We shall now show that the intensity fluctuations $\delta \hat I_{\pm\bm k,\mathrm{out}}$ of thus obtained "signal" and "idler" are perfectly correlated, \textit{i. e.}, they represent the so-called twin beams \cite{Klyshko1967, Heidmann1987}.

To this end, we substitute $\hat{\mathsf b}_{ \pm\bm k}=\mathsf b_{ \pm\bm k}+\delta \hat{\mathsf b}_{\pm \bm k}$ into Eq. \eqref{HLequation} and account for the vacuum fluctuations at the edges:
\begin{equation*}
\begin{split}
0&=-i\bar \Delta_{\bm k}\delta \hat{\mathsf b}_{\bm k}-\frac{\kappa}{2}\delta \hat{\mathsf b}_{\bm k}-\sqrt{\kappa}\delta \hat{\mathsf e}_{\bm k,\mathrm{in}}\\
&-\frac{i}{\hbar}g\left[2\delta\hat n_0\mathsf b_{\bm k}+2n_0 \delta \hat{\mathsf b}_{\bm k}+\tfrac{1}{S}\mathsf b_{0}^2\delta \hat{\mathsf b}_{-\bm k}^\dagger+\tfrac{2}{S} \mathsf b_{0}\hat{\mathsf b}_{-\bm k}^\dagger\delta \hat{\mathsf b}_{0}\right]
\end{split}
\end{equation*}
where $\delta\hat n_0=\delta\hat n_0^\dagger =\tfrac{1}{S}(\mathsf b_{0}^\ast \delta \hat{\mathsf b}_{0}+\mathsf b_{0}\delta \hat{\mathsf b}_{0}^\dagger)$, and we  have defined $\bar\Delta_{\bm k}\equiv \omega_\uparrow(\bm k)-\Delta_{\uparrow\uparrow}(\bm k)-\omega$. Next, we combine the above set into corresponding equations for the rotated quadratures
\begin{equation*}
\begin{split}
\hat x_{\pm\bm k}&= \tfrac{1}{2}\left (\hat{\mathsf b}_{ \pm\bm k}e^{-i\phi_{\pm\bm k}}+\hat{\mathsf b}_{ \pm\bm k}^\dagger e^{i\phi_{\pm\bm k}}\right)\\ 
\hat y_{\pm\bm k}&= \tfrac{1}{2i}\left (\hat{\mathsf b}_{ \pm\bm k}e^{-i\phi_{\pm\bm k}}-\hat{\mathsf b}_{ \pm\bm k}^\dagger e^{i\phi_{\pm\bm k}}\right)
\end{split}
\end{equation*}
and use Eq. \eqref{PhaseRelation} for the phases. Subtract one of thus obtained equations for the quadrature fluctuations from the other:
\begin{equation*}
\begin{split}
&-\bar\Delta_{\bm k}(\delta \hat y_{\bm k}-\delta \hat y_{-\bm k})+\frac{\kappa}{2}(\delta \hat x_{\bm k}-\delta \hat x_{-\bm k})=-\frac{gn_0}{\hbar}(\delta \hat x_{\bm k}-\delta \hat x_{-\bm k})\\
&+2\frac{gn_0}{\hbar}(\delta \hat y_{\bm k}-\delta \hat y_{-\bm k})-\sqrt{\kappa}(\delta \hat x_{\bm k,\mathrm{in}}-\delta \hat x_{-\bm k,\mathrm{in}}).
\end{split}
\end{equation*}
By virtue of Eq. \eqref{ThresholdDensity} and Eq. \eqref{OptimalDetuning} this relation may further be reduced to
\begin{equation*}
-\sqrt{\kappa}(\delta \hat x_{\bm k}-\delta \hat x_{-\bm k})=(\delta \hat x_{\bm k,\mathrm{in}}-\delta \hat x_{-\bm k,\mathrm{in}}).
\end{equation*}
Furthermore, by using the limiting form of the input-output relation \eqref{BoundaryCond1} for oblique incidence
\begin{equation}
\label{InputOutputEdges}
\hat e_{\pm\bm k,\mathrm{out}}-\hat e_{\pm \bm k,\mathrm{in}}=\sqrt{\kappa}\hat b_{\pm\bm k},
\end{equation}
we obtain $\delta \hat x_{\bm k,\mathrm{out}}-\delta \hat x_{-\bm k,\mathrm{out}}=0$, which implies
\begin{equation}
\delta \hat I_{\bm k,\mathrm{out}}-\delta \hat I_{-\bm k,\mathrm{out}}\equiv 2\sqrt{\kappa n_{\bm k}}(\delta \hat x_{\bm k,\mathrm{out}}-\delta \hat x_{-\bm k,\mathrm{out}})=0
\end{equation}
with $n_{\bm k}$ being the polariton beam density.

Finally, let us assume $n_0<n_\mathrm{th}$, so that no instability occurs and one may accumulate the excitons at $\bm k=0$ by increasing the pump power. A monochromatic polariton wave $e_{-\bm k,\mathrm{in}}(t)=\mathsf e_{\mathrm{in}}e^{-i\omega t}$ having the frequency $\omega$ identical to that of the pump impinges onto one (pair) of the edges (see Fig. \ref{Fig1}). By using the input-output relation \eqref{InputOutputEdges}, one readily obtains
\begin{equation}
\label{PhaseConjugation}
\hat {\mathsf e}_{\bm k,\mathrm{out}}=-\frac{i g \mathsf b_{0}^2\kappa}{\hbar [(\varepsilon_\uparrow(\bm k)-\omega)^2+\tfrac{\kappa^2}{4}-(gn_0/\hbar)^2]}\hat {\mathsf e}_{-\bm k,\mathrm{in}}^\dagger,
\end{equation}
which relates the input and output signal at the same edge.\footnote{Phase conjugation persists for partially transmitting boundaries. In a high-finesse cavity, however, the conjugate wave leaves at resonance through both edges, its amplitude at each of them being half of that in Eq. \eqref{PhaseConjugation}, and at the input edge it is accompanied by a directly reflected probe wave.} In order to reveal the physical meaning of this relation, we recall Eq. \eqref{Polarization} for the exciton polarization operator and evaluate its expectation value on the initial (input) coherent state:
\begin{equation}
\label{ConjugatedWave}
\bm P_{\uparrow,\mathrm{out}}(\bm r,t)\sim \frac{|\mathsf e_{\mathrm{in}}|e^{i(\bm k\bm r-\omega t)}e^{-i\phi_\mathrm{in}}}{ [\varepsilon_\uparrow(\bm k)-\omega]^2+\tfrac{\kappa^2}{4}-(gn_0/\hbar)^2}+c.c.
\end{equation}
On the other hand, the input polarization reads
\begin{equation*}
\bm P_{\uparrow,\mathrm{in}}(\bm r,t)\sim |\mathsf e_{\mathrm{in}}|e^{i(-\bm k\bm r-\omega t)}e^{i\phi_\mathrm{in}}+c.c.
\end{equation*}        
It is evident that, up to a constant phase that can be adjusted by the pump, the reflected polarization wave is phase-conjugated to the input one: it represents the input wave propagating "backward in time" \cite{Zeldovich1985}. Eq. \eqref{ConjugatedWave} shows that the phase-conjugated wave can be significantly amplified when the denominator on the r.h.s. approaches zero. This corresponds to the threshold for the modulational instability discussed previously.

\subsection{Ultra-fast longitudinal polariton beams}

Having established the basic physics in the single-component limit, we now briefly address the full two-component problem at zero magnetic field, $B_z=0$, the regime most directly relevant to the recent experiments on exciton mirrors \cite{Back2018, Scuri2018}. As a first approximation, we set $\Delta_{\uparrow\downarrow} = \Delta_{\downarrow\uparrow}^{\ast} = 0$ inside the light cone, which therefore may still be defined as \(k<k_\uparrow=k_\downarrow\). The bistable response to a circularly polarized pump at normal incidence, \(\bm k=0\),  then remains governed by Eq. \eqref{Bistability}. For \(k>k_\uparrow\) the polariton spectrum comprises a longitudinal (\(p\)-polarized), and a transverse (\(s\)-polarized) branches,  \(\varepsilon_p(\bm k)\) and \(\varepsilon_s(\bm k)\), respectively. These are obtained by diagonalizing Eqs. \eqref{ExcHeisenbergLangevinEq}, with \(\Delta_{\sigma\sigma^\prime}\) given by Eq. \eqref{EnergyShifts}. For $\kappa<\gamma_0$ and sufficiently negative detuning, $\bar\Delta_0\leqslant \bar\Delta_0^{(c)}\equiv -\hbar k_\uparrow^2/2m-(2+g_{\uparrow\downarrow}/g)\kappa/2$, the circularly-polarized bistability is preempted by an instability into the \(p\)-polarized polariton modes at opposite momenta \(\pm\bm k_\ast\), where \(k_\ast\) is determined by
\begin{equation}
\label{LongitudinalResonanceCondition}
\begin{split}
\omega&=\varepsilon_{p}(\bm k_\ast)\\
&= \omega_\uparrow (\bm k_\ast)+\frac{\upsilon_{\uparrow\downarrow}}{c}\sqrt{k_\ast^2 c^2-\varepsilon_{p}^2(\bm k_\ast)}+\frac{2g+g_{\uparrow\downarrow}}{2\hbar}n_0,
\end{split}
\end{equation}
with the corresponding threshold density being twice that found in the spin-polarized case, \textit{i.e.}, $n_0=2n_\mathrm{th}$. For $\bar\Delta_0^{(c)}-\bar\Delta_0\lesssim \upsilon_{\uparrow\downarrow} k_p$, one obtains
\begin{equation}
\label{GroupVelocity}
\frac{\partial\varepsilon_p(\bm k_\ast)}{\partial k}=c\frac{\sqrt{1+\left (\frac{\bar\Delta_0^{(c)}-\bar\Delta_0}{\upsilon_{\uparrow\downarrow} k_p}\right)^2}}{1+\frac{c}{\upsilon_{\uparrow\downarrow}}\left (\frac{\bar\Delta_0^{(c)}-\bar\Delta_0}{\upsilon_{\uparrow\downarrow} k_p}\right)},
\end{equation}
showing explicitly that the velocity of polariton beams can approach \(c\). 

Intriguingly, the recent experiment \cite{Aguila2023} has revealed ultrafast collective propagation of 2D excitons under near-resonant excitation at normal incidence and above a threshold density. The possible connection between this phenomenon and the polariton instability of the exciton mirror predicted here will be investigated in a separate work within a more comprehensive treatment of the two-component model, including the regime $\bar\Delta_0> \bar\Delta_0^{(c)}$.

\section{Conclusions}

In conclusion, our analysis reveals that the dispersive optical bistability anticipated for a two-dimensional exciton gas subjected to a coherent drive within the light cone is susceptible to a modulational instability toward non-radiative polariton modes. The predicted phenomenology of ultra-fast twin polariton beams and phase conjugation is inherent to ensembles of bright excitons and emerges at the most fundamental level of the theory, suggesting a new intriguing direction for the development of the semiconductor quantum optics.

\acknowledgements

The Appendices for this work have been written in collaboration with Chat GPT-6 Astra Pro. The Chat GPT output has been revised by the author. Responsibility for the final manuscript lies entirely with the author. GAI tool is not listed as author and does not bear responsibility for the final outcomes.

We also thank M. A. Semina for pointing out that, for a monolayer, the nonradiative modes can be referred to simply as polaritons, without the qualifier “surface.” Although these modes share important analogies with surface modes of bulk media, we agree that this qualifier could be misleading in the present context.     

\appendix

\section{Representative values of the parameters}

Representative values of the parameters are provided below. In most cases, we find a good agreement of our analytic theory with the relevant experiments and \textit{ab-initio} calculations.    

\label{sec:parameters}

\subsection{MoSe$_2$ exciton mirrors}

\begin{table}[t]
\caption{Representative parameters for the MoSe$_2$ $A$ exciton.
The density estimates use $g=0.1~\mu{\rm eV}\,\mu{\rm m}^{2}$;
$n_{\rm th}$ additionally assumes the illustrative choice
$\kappa=\gamma_0/2$.}
\label{tab:parameters}
\begin{ruledtabular}
\begin{tabular}{lc}
Quantity & Representative value \\
$\hbar\omega_0$ & $1.64~{\rm eV}$ \\
$q_0=\omega_0/c$ & $8.31~\mu{\rm m}^{-1}$ \\
$m$ & $\simeq1.1m_e$ \\
$a_X$ & $\simeq1~{\rm nm}$ \\
$|p_{cv}|/m_e$ & $\simeq4.7\times10^5~{\rm m/s}$ \\
$\hbar\gamma_0$ & $\simeq1.7~{\rm meV}$ \\
$\upsilon_{\uparrow\uparrow}/c$ & $\simeq1.0\times10^{-3}$ \\
$\frac{\hbar q_0^2}{2m\omega_0}$ & $\simeq1.5\times10^{-6}$ \\
$g_{\sigma\sigma'}$ & $\sim0.1$--$0.14~\mu{\rm eV}\,\mu{\rm m}^{2}$ \\
$\kappa$ (illustrative) & $\gamma_0/2$ \\
$n_{\min}^{(-)}$ & $\simeq1.7\times10^4~\mu{\rm m}^{-2}$ \\
$n_{\min}^{(+)}$ & $\simeq2.8\times10^4~\mu{\rm m}^{-2}$ \\
$n_{\rm th}$ & $\simeq4.3\times10^3~\mu{\rm m}^{-2}$ \\
$|\mathsf{g}|$ & $\simeq4$ \\
$|B_z|$ ($k\simeq q_0$) & $\gtrsim 7.3~{\rm T}$ \\
\end{tabular}
\end{ruledtabular}
\end{table}

For a representative MoSe$_2$ $A$-exciton resonance
$\hbar\omega_0\simeq1.64~{\rm eV}$, one has
$q_0=\omega_0/c\simeq8.31~\mu{\rm m}^{-1}$. We use
$m\simeq1.1m_e$ and $a_X\simeq1$~nm, the latter being the
characteristic exciton radius quoted for MoSe$_2$ in
Ref.~\cite{Scuri2018}. The standard massive-Dirac parameters
$a_0\simeq3.31$~\AA{} and $t\simeq0.94$~eV \cite{Xiao2012} yield the interband
velocity
$|p_{cv}|/m_e\simeq a_0t/\hbar\simeq4.7\times10^5~{\rm m/s}$.
Here $p_{cv}=\langle c|p_x|v\rangle$ denotes a single Cartesian
component of the interband momentum matrix element.
Thus, Eq.~(16) of the main text, with
$|\psi_{1s}(0)|^2\simeq2/(\pi a_X^2)$, yields
\begin{equation*}
 \gamma_0\simeq
 \frac{e^2(|p_{cv}|/m_e)^2}{2c\varepsilon_0\hbar\omega_0}
 \frac{2}{\pi a_X^2},
 \qquad
 \hbar\gamma_0\simeq1.7~{\rm meV}.
 \label{eq:gamma0_parameters}
\end{equation*}
In the linewidth convention of Ref.~\cite{Scuri2018}, this
corresponds to the full radiative linewidth
$2\hbar\gamma_0\simeq3.4$~meV, which is within the experimentally
extracted $2$--$4$~meV range. The same parameters give
\begin{equation*}
 \upsilon_{\uparrow\uparrow}
 =\frac{\gamma_0}{q_0}
 =\frac{c\gamma_0}{\omega_0}
 \simeq1.0\times10^{-3}c,
 \qquad
 \frac{\hbar q_0^2}{2m\omega_0}
 \simeq1.5\times10^{-6}.
 \label{eq:vbeta_parameters}
\end{equation*}
Hence, one has $k_\uparrow\simeq q_0$. Beyond the narrow
region near the light cone the term linear in $k$
dominates the bare parabolic contribution over the frequency range
of interest.

The interaction constants may be estimated from the low-energy 2D
scattering amplitude used in the main text,
\begin{equation}
 g_{\sigma\sigma'}=-\frac{4\pi\hbar^2}{m}
 {\rm Re}\!\left[
 \frac{1}{\ln(na_{\sigma\sigma'}^2)-i\pi}
 \right].
 \label{eq:g_parameters}
\end{equation}
For $a_{\sigma\sigma'}\sim1$--$3$~nm and
$n\sim10^3$--$10^4~\mu{\rm m}^{-2}$,
Eq.~(\ref{eq:g_parameters}) gives
$g_{\sigma\sigma'}\sim0.1$--$0.14~
\mu{\rm eV}\,\mu{\rm m}^{2}$. This scale agrees with
the nonlinear blueshift coefficient extracted in
Ref.~\cite{Scuri2018},
$0.096~\mu{\rm eV}\,\mu{\rm m}^{2}$.
Taking $g\simeq0.1~\mu{\rm eV}\,\mu{\rm m}^{2}$, the two
turning-point densities at the optimal detuning
$\bar\Delta_0=-2\gamma_0$ are
$n_{\min}^{(-)}=\hbar\gamma_0/g\simeq1.7\times10^4~\mu{\rm m}^{-2}$ and
$n_{\min}^{(+)}=5\hbar\gamma_0/(3g)\simeq2.8\times10^4~\mu{\rm m}^{-2}$.
Here the upper turning point is evaluated at the same detuning that
minimizes the lower turning-point density.
For the choice $\kappa=\gamma_0/2$ used for illustrative purposes in Fig. \ref{Fig1},
$n_{\rm th}=\hbar\gamma_0/(4g)
\simeq4.3\times10^3~\mu{\rm m}^{-2}$.
Besides Ref.~\cite{Scuri2018}, recent nanocavity experiments also demonstrate low-density optical non-linearities \cite{Wang2026}, although a quantitative density comparison
requires conversion of the reported switching energy into an exciton
areal density. In practice, we expect $\kappa\ll\gamma_0$ for the exchange-dominated
linear part of the polariton dispersion within the boundary-leakage
model considered below.

 \subsection{Magnetic fields required to reach the single-component limit and the biexciton Feshbach resonance}

For the MoSe$_2$ bright exciton one has $|\mathsf{g}|\simeq4$ \cite{Stier2016}.
The single-component limit requires
$\mu_B|\mathsf{g}B_z|\gg
\hbar\upsilon_{\uparrow\downarrow}k$.
If
$\upsilon_{\uparrow\downarrow}\sim\upsilon_{\uparrow\uparrow}$,
the Zeeman and exchange scales evaluated at $k=q_0$ become equal for
\begin{equation*}
 |B_z|
 \sim
 \frac{\hbar\upsilon_{\uparrow\downarrow}q_0}
 {\mu_B|\mathsf{g}|}\sim
 \frac{\hbar\gamma_0}{\mu_B|\mathsf{g}|}
 \simeq7.3~{\rm T}.
 \label{eq:B_parameters}
\end{equation*}
Thus fields of several teslas bring the two energy scales into
competition near the light cone, while a single-component limit requires $|B_z|$ appreciably exceeding this
crossover value, typically fields of order a few tens of teslas near
$k=q_0$. At larger momenta, the crossover field increases approximately
as $k/q_0$. This range overlaps with the field scale discussed for the
biexciton Feshbach resonance in Ref. \cite{Andreev2024}; the actual
resonance field depends on the material parameters and the biexciton
binding energy.  

\subsection{MoS$_2$ and the high-density two-component regime}

It is also useful to give analogous estimates for MoS$_2$, since the
full two-component problem at $B_z=0$ may be relevant to the ultrafast
exciton-flow experiment of Ref.~\cite{Aguila2023}. Representative
parameters are collected in Table~\ref{tab:parameters_MoS2}.

\begin{table}[t]
\caption{Representative parameters for the MoS$_2$ $A$ exciton and
the scales relevant to Ref.~\cite{Aguila2023}.}
\label{tab:parameters_MoS2}
\begin{ruledtabular}
\begin{tabular}{lc}
Quantity & Representative value \\
$\hbar\omega_0$ & $1.95~{\rm eV}$ \\
$q_0=\omega_0/c$ & $9.88~\mu{\rm m}^{-1}$ \\
$m$ & $\simeq0.8m_e$ \\
$a_X$ & $\simeq1.3~{\rm nm}$ \\
$|p_{cv}|/m_e$ & $\simeq5.3\times10^5~{\rm m/s}$ \\
$\hbar\gamma_0$ & $\simeq1.1~{\rm meV}$ \\
$\upsilon_{\sigma\sigma'}/c$ & $\sim5.6\times10^{-4}$ \\
$\frac{\hbar q_0^2}{2m\omega_0}$ & $\simeq2.4\times10^{-6}$ \\
$g_{\sigma\sigma}$ & $\sim0.1$--$0.2~\mu{\rm eV}\,\mu{\rm m}^{2}$ \\
$n$ & $\sim2\times10^5~\mu{\rm m}^{-2}$ \\
$v$ & $1.8\times10^7~{\rm m/s}\simeq0.06c$
\end{tabular}
\end{ruledtabular}
\end{table}

Using the neutral-exciton energy
$\hbar\omega_0\simeq1.95~{\rm eV}$ reported in
Ref.~\cite{Aguila2023}, together with the standard massive-Dirac
parameters $a_0\simeq3.19$~\AA{} and $t\simeq1.10$~eV \cite{Xiao2012}, gives
\begin{equation*}
    \frac{|p_{cv}|}{m_e}
    \simeq\frac{a_0t}{\hbar}
    \simeq5.3\times10^5~{\rm m/s}.
\end{equation*}
Taking $a_X\simeq1.3$~nm and $m\simeq0.8m_e$, Eq.~(16) of the
main text yields
\begin{equation*}
    \hbar\gamma_0\simeq1.1~{\rm meV},
    \qquad
    \frac{\upsilon_{\sigma\sigma'}}{c}
    \sim\frac{\gamma_0}{\omega_0}
    \simeq5.6\times10^{-4},
    \label{eq:MoS2_radiative}
\end{equation*}
where in the last estimate we assume identical oscillator strengths
for the two bright components. This corresponds to
$\upsilon_{\sigma\sigma'}\sim1.7\times10^5~{\rm m/s}$, of
the same order of magnitude as the exchange-induced longitudinal
velocity $\sim3\times10^5~{\rm m/s}$ obtained from first-principles
calculations \cite{Qiu2015}. One also finds $\hbar q_0^2/2m\omega_0\simeq2.4\times10^{-6}$.
Thus, as for MoSe$_2$, one has $k_\sigma\simeq q_0$, and the polariton dispersion beyond the light cone is dominated by the linear term due to the long-range
exchange.

It is instructive to compare our interaction estimates with
Ref.~\cite{Aguila2023}. In the pump--probe measurements the reported
estimate of the exciton density under the excitation spot reaches
$n\lesssim2.4\times10^{13}~{\rm cm}^{-2}
=2.4\times10^5~\mu{\rm m}^{-2}$.
For a representative density $n\simeq2\times10^5~\mu{\rm m}^{-2}$
and $a_{\sigma\sigma}\sim1$~nm, Eq.~(\ref{eq:g_parameters})
gives
\begin{equation*}
    g_{\sigma\sigma}
    \simeq0.15~
    \mu{\rm eV}\,\mu{\rm m}^{2},
    \qquad
    g_{\sigma\sigma}n\sim30~{\rm meV}.
    \label{eq:MoS2_interaction}
\end{equation*}
The latter scale exceeds by more than an order of magnitude the
$\sim2$~meV exciton blueshift observed in
Ref.~\cite{Aguila2023}. This comparison concerns characteristic
scales: the quoted density is the injected density under the pump,
whereas the measured blueshift varies with position and time as the
exciton population expands. At such elevated densities, moreover,
the gas parameter is $na_{\sigma\sigma}^2\simeq0.2$ for
$a_{\sigma\sigma}=1$~nm and is no longer asymptotically small; the
dilute-limit two-body description, particularly in the opposite-spin
channel, need not remain quantitatively applicable. Both short-range attraction due to the exchange of identical fermions and van der Waals forces may then substantially renormalize the effective
$g_{\uparrow\downarrow}$. It is therefore natural, at a phenomenological
level, to allow for $g_{\uparrow\downarrow}<0$.\footnote{Here and in the dilute limit assumed in the main text, \textit{inelastic} scattering into the biexcitons has been neglected. Moreover, at the densities reported in \cite{Aguila2023}, the average distance between the excitons approaches the typical biexciton radius, so that biexcitons may actually never form in such conditions (the so-called "biexciton Mott transition").} According to Eq. \eqref{LongitudinalResonanceCondition}
of the main text, the
interaction-induced shift of the (longitudinal) polariton dispersion $\varepsilon_p(\bm k)$ is then controlled by the combination
$g_{\uparrow\uparrow}+g_{\uparrow\downarrow}/2$, which may be
considerably smaller than $g_{\uparrow\uparrow}$ alone.

The experiment reports propagation of the exciton fluid with
$v\simeq1.8\times10^7~{\rm m/s}\simeq0.06c$ well beyond the pump region \cite{Aguila2023}. This velocity is substantially larger than $\upsilon_{\uparrow\downarrow}$ and should therefore not be identified with the bare exchange scale. Within the present polariton interpretation, such a velocity would
require the characteristic momentum of the instability $\bm k_\ast$
to approach the light cone, where retardation becomes important. Indeed, our Eq. \eqref{GroupVelocity} in the main text shows that the group velocity of the $p$-polarized polariton beams may approach $c$ when $\bar\Delta_0$  approaches the critical detuning $\bar\Delta_0^{(c)}$. The latter corresponds to the frequency at which the blue-shifted longitudinal-polariton dispersion $\varepsilon_p(\bm k)$ enters the light cone. For the boundary parameters used below, the polariton-decay model gives $\hbar\kappa\approx1.2$~meV at the reported experimental velocity [Eq.~(B2)]. Thus, for interaction ratios of order unity, the blueshift at the threshold density inferred from this model is of the order of the experimental meV scale.

\section{Estimate of the polariton decay rate within the adopted model}
\label{sec:kappa_estimate}

It is instructive to obtain an estimate of the polariton decay rate $\kappa$ within our model, \textit{i.e.}, by considering a square monolayer domain of area
$S$, separated by a partially transmitting boundary from an
otherwise infinite monolayer. Nanometre-scale gaps have been
shown to provide controllable transmission of two-dimensional guided
modes in graphene \cite{Nikitin2013}. We use this observation only
as a proof of principle and characterize the boundary by an
amplitude transmission coefficient $t$. Neglecting out-of-plane scattering at
the boundary, the intensity reflection coefficient is then
$R=1-|t|^2$.

Consider a polariton beam with group velocity $\bm v(k)$ propagating at an
angle $\theta$ with respect to one of the sides of the square. The
rates at which it encounters the two pairs of opposite boundaries
are $v(k)|\cos\theta|/\sqrt{S}$ and $v(k)|\sin\theta|/\sqrt{S}$, respectively.
Consequently, its occupation decays at the rate
\begin{equation*}
 \kappa_{\bm k}
 =
 -\frac{v(k)}{\sqrt{S}}
 \left(
 |\cos\theta|+|\sin\theta|
 \right)
 \ln\left(1-|t|^2\right).
 \label{eq:kappa_finite_sample}
\end{equation*}
Here $\kappa_{\bm k}$ is the occupation-decay rate; the corresponding
amplitude-decay rate is $\kappa_{\bm k}/2$.
For direction-independent gain and boundary transmission, this
leakage favors $\theta=0$ ($\mathrm{mod}$ $\pi/2$), which minimizes
the boundary-collision rate. For simplicity, we may take $\theta=0$ as a representative value also for the phase-conjugation phenomenon. We obtain
\begin{equation}
 \kappa
 \simeq
 -\frac{v(k)}{\sqrt{S}}
 \ln\left(1-|t|^2\right)
 \simeq
 |t|^2\frac{v(k)}{\sqrt{S}},
 \qquad |t|^2\ll1.
 \label{eq:kappa_average}
\end{equation}

In the momentum range where the exchange-induced linear
contribution to the polariton dispersion dominates, one
has $v\sim \upsilon_{\uparrow\uparrow}$. Using the typical value obtained above,
$\upsilon_{\uparrow\uparrow}/c\sim 10^{-3}$, gives
$v\sim3\times10^5~{\rm m/s}$. Large-area monolayers with $S\simeq 10^3~\mu{\rm m}^2$ and lateral
dimensions of several tens of micrometres have been demonstrated
experimentally \cite{Aguila2023}. Replacing such a flake by an
equal-area square gives $\sqrt{S}\simeq31.6~\mu{\rm m}$. For the typical
value $|t|\sim 0.1$, Eq.~(\ref{eq:kappa_average}) yields
\begin{equation*}
 \kappa^{-1}\sim 10~{\rm ns},
 \qquad
 \hbar\kappa
 \sim6\times10^{-2}~\mu{\rm eV}.
\end{equation*}
Thus, for $\hbar\gamma_0\simeq1.7~{\rm meV}$, one finds
$\kappa/\gamma_0\sim4\times10^{-5}$. Edge leakage from a
large monolayer enclosed by reflecting boundaries
therefore gives the regime $\kappa\ll2\gamma_0$
assumed in the main text within this idealized model.

For fully transparent boundaries, $|t|\to1$, the pumped region acts as a single-pass parametric amplifier with counter-propagating signal and idler, i.e., a mirrorless parametric oscillator~\cite{Harris1966, Canalias2007}. Within the phenomenological extension of the Heisenberg-Langevin equation~\eqref{HLequation}, the stationary edge fields are then reproduced exactly by setting
\begin{equation}
\hbar\kappa=2gn_0\cot\left[\frac{gn_0\sqrt{S}}{2\hbar v(k)}\right],
\end{equation}
where $\sqrt{S}$ is now the size of the region occupied by the mean-field background. At the threshold, $n_0=n_\mathrm{th}$, this gives $\kappa=\pi v(k)/\sqrt{S}$, the inverse time of flight across this region. For $\sqrt{S}\simeq31.6~\mu$m and the exchange-dominated part of the dispersion, $v\sim3\times10^5$~m/s, one finds $\hbar\kappa\approx20~\mu$eV, still much smaller than $\hbar\gamma_0$. For the velocity $v\simeq0.06c$ reported in Ref.~\cite{Aguila2023}, however, one gets $\hbar\kappa\approx1.2$~meV, so that the two decay rates become comparable.

\bibliography{Bibliography}

\begin{thebibliography}{49}%
\makeatletter
\providecommand \@ifxundefined [1]{%
 \@ifx{#1\undefined}
}%
\providecommand \@ifnum [1]{%
 \ifnum #1\expandafter \@firstoftwo
 \else \expandafter \@secondoftwo
 \fi
}%
\providecommand \@ifx [1]{%
 \ifx #1\expandafter \@firstoftwo
 \else \expandafter \@secondoftwo
 \fi
}%
\providecommand \natexlab [1]{#1}%
\providecommand \enquote  [1]{``#1''}%
\providecommand \bibnamefont  [1]{#1}%
\providecommand \bibfnamefont [1]{#1}%
\providecommand \citenamefont [1]{#1}%
\providecommand \href@noop [0]{\@secondoftwo}%
\providecommand \href [0]{\begingroup \@sanitize@url \@href}%
\providecommand \@href[1]{\@@startlink{#1}\@@href}%
\providecommand \@@href[1]{\endgroup#1\@@endlink}%
\providecommand \@sanitize@url [0]{\catcode `\\12\catcode `\$12\catcode
  `\&12\catcode `\#12\catcode `\^12\catcode `\_12\catcode `\%12\relax}%
\providecommand \@@startlink[1]{}%
\providecommand \@@endlink[0]{}%
\providecommand \url  [0]{\begingroup\@sanitize@url \@url }%
\providecommand \@url [1]{\endgroup\@href {#1}{\urlprefix }}%
\providecommand \urlprefix  [0]{URL }%
\providecommand \Eprint [0]{\href }%
\providecommand \doibase [0]{https://doi.org/}%
\providecommand \selectlanguage [0]{\@gobble}%
\providecommand \bibinfo  [0]{\@secondoftwo}%
\providecommand \bibfield  [0]{\@secondoftwo}%
\providecommand \translation [1]{[#1]}%
\providecommand \BibitemOpen [0]{}%
\providecommand \bibitemStop [0]{}%
\providecommand \bibitemNoStop [0]{.\EOS\space}%
\providecommand \EOS [0]{\spacefactor3000\relax}%
\providecommand \BibitemShut  [1]{\csname bibitem#1\endcsname}%
\let\auto@bib@innerbib\@empty
\bibitem [{\citenamefont {Boyd}(2020)}]{Boyd}%
  \BibitemOpen
  \bibfield  {author} {\bibinfo {author} {\bibfnamefont {R.~W.}\ \bibnamefont
  {Boyd}},\ }\href@noop {} {\emph {\bibinfo {title} {Nonlinear optics}}},\
  \bibinfo {edition} {4th}\ ed.\ (\bibinfo  {publisher} {Elsevier Academic
  Press},\ \bibinfo {year} {2020})\BibitemShut {NoStop}%
\bibitem [{\citenamefont {Ivchenko}(2004)}]{Ivchenko2004}%
  \BibitemOpen
  \bibfield  {author} {\bibinfo {author} {\bibfnamefont {E.~L.}\ \bibnamefont
  {Ivchenko}},\ }\href@noop {} {\emph {\bibinfo {title} {Optical Spectroscopy
  of Semiconductor Nanostructures}}}\ (\bibinfo  {publisher} {Springer Berlin,
  Heidelberg},\ \bibinfo {year} {2004})\BibitemShut {NoStop}%
\bibitem [{\citenamefont {Wild}\ \emph {et~al.}(2018)\citenamefont {Wild},
  \citenamefont {Shahmoon}, \citenamefont {Yelin},\ and\ \citenamefont
  {Lukin}}]{Wild2018}%
  \BibitemOpen
  \bibfield  {author} {\bibinfo {author} {\bibfnamefont {D.~S.}\ \bibnamefont
  {Wild}}, \bibinfo {author} {\bibfnamefont {E.}~\bibnamefont {Shahmoon}},
  \bibinfo {author} {\bibfnamefont {S.~F.}\ \bibnamefont {Yelin}},\ and\
  \bibinfo {author} {\bibfnamefont {M.~D.}\ \bibnamefont {Lukin}},\ }\bibfield
  {title} {\bibinfo {title} {Quantum nonlinear optics in atomically thin
  materials},\ }\href {https://doi.org/10.1103/PhysRevLett.121.123606}
  {\bibfield  {journal} {\bibinfo  {journal} {Phys. Rev. Lett.}\ }\textbf
  {\bibinfo {volume} {121}},\ \bibinfo {pages} {123606} (\bibinfo {year}
  {2018})}\BibitemShut {NoStop}%
\bibitem [{\citenamefont {Back}\ \emph {et~al.}(2018)\citenamefont {Back},
  \citenamefont {Zeytinoglu}, \citenamefont {Ijaz}, \citenamefont {Kroner},\
  and\ \citenamefont {Imamo\ifmmode~\breve{g}\else \u{g}\fi{}lu}}]{Back2018}%
  \BibitemOpen
  \bibfield  {author} {\bibinfo {author} {\bibfnamefont {P.}~\bibnamefont
  {Back}}, \bibinfo {author} {\bibfnamefont {S.}~\bibnamefont {Zeytinoglu}},
  \bibinfo {author} {\bibfnamefont {A.}~\bibnamefont {Ijaz}}, \bibinfo {author}
  {\bibfnamefont {M.}~\bibnamefont {Kroner}},\ and\ \bibinfo {author}
  {\bibfnamefont {A.}~\bibnamefont {Imamo\ifmmode~\breve{g}\else
  \u{g}\fi{}lu}},\ }\bibfield  {title} {\bibinfo {title} {Realization of an
  electrically tunable narrow-bandwidth atomically thin mirror using monolayer
  ${\mathrm{mose}}_{2}$},\ }\href
  {https://doi.org/10.1103/PhysRevLett.120.037401} {\bibfield  {journal}
  {\bibinfo  {journal} {Phys. Rev. Lett.}\ }\textbf {\bibinfo {volume} {120}},\
  \bibinfo {pages} {037401} (\bibinfo {year} {2018})}\BibitemShut {NoStop}%
\bibitem [{\citenamefont {Scuri}\ \emph {et~al.}(2018)\citenamefont {Scuri},
  \citenamefont {Zhou}, \citenamefont {High}, \citenamefont {Wild},
  \citenamefont {Shu}, \citenamefont {De~Greve}, \citenamefont {Jauregui},
  \citenamefont {Taniguchi}, \citenamefont {Watanabe}, \citenamefont {Kim},
  \citenamefont {Lukin},\ and\ \citenamefont {Park}}]{Scuri2018}%
  \BibitemOpen
  \bibfield  {author} {\bibinfo {author} {\bibfnamefont {G.}~\bibnamefont
  {Scuri}}, \bibinfo {author} {\bibfnamefont {Y.}~\bibnamefont {Zhou}},
  \bibinfo {author} {\bibfnamefont {A.~A.}\ \bibnamefont {High}}, \bibinfo
  {author} {\bibfnamefont {D.~S.}\ \bibnamefont {Wild}}, \bibinfo {author}
  {\bibfnamefont {C.}~\bibnamefont {Shu}}, \bibinfo {author} {\bibfnamefont
  {K.}~\bibnamefont {De~Greve}}, \bibinfo {author} {\bibfnamefont {L.~A.}\
  \bibnamefont {Jauregui}}, \bibinfo {author} {\bibfnamefont {T.}~\bibnamefont
  {Taniguchi}}, \bibinfo {author} {\bibfnamefont {K.}~\bibnamefont {Watanabe}},
  \bibinfo {author} {\bibfnamefont {P.}~\bibnamefont {Kim}}, \bibinfo {author}
  {\bibfnamefont {M.~D.}\ \bibnamefont {Lukin}},\ and\ \bibinfo {author}
  {\bibfnamefont {H.}~\bibnamefont {Park}},\ }\bibfield  {title} {\bibinfo
  {title} {Large excitonic reflectivity of monolayer ${\mathrm{mose}}_{2}$
  encapsulated in hexagonal boron nitride},\ }\href
  {https://doi.org/10.1103/PhysRevLett.120.037402} {\bibfield  {journal}
  {\bibinfo  {journal} {Phys. Rev. Lett.}\ }\textbf {\bibinfo {volume} {120}},\
  \bibinfo {pages} {037402} (\bibinfo {year} {2018})}\BibitemShut {NoStop}%
\bibitem [{\citenamefont {Agranovich}\ and\ \citenamefont
  {Ginzburg}(1984)}]{Agranovitch}%
  \BibitemOpen
  \bibfield  {author} {\bibinfo {author} {\bibfnamefont {V.~M.}\ \bibnamefont
  {Agranovich}}\ and\ \bibinfo {author} {\bibfnamefont {V.}~\bibnamefont
  {Ginzburg}},\ }\href@noop {} {\emph {\bibinfo {title} {Crystal optics with
  spatial dispersion, and excitons}}},\ \bibinfo {series} {Springer Series in
  Solid-State Sciences}, Vol.~\bibinfo {volume} {42}\ (\bibinfo  {publisher}
  {Springer Berlin Heidelberg},\ \bibinfo {address}
  {https://doi.org/10.1007/978-3-662-02406-5},\ \bibinfo {year}
  {1984})\BibitemShut {NoStop}%
\bibitem [{\citenamefont {Gibbs}(1985)}]{Gibbs}%
  \BibitemOpen
  \bibfield  {author} {\bibinfo {author} {\bibfnamefont {H.~M.}\ \bibnamefont
  {Gibbs}},\ }\href@noop {} {\emph {\bibinfo {title} {Optical Bistability:
  Controlling Light with Light}}}\ (\bibinfo  {publisher} {Academic Press},\
  \bibinfo {year} {1985})\BibitemShut {NoStop}%
\bibitem [{\citenamefont {Scully}\ and\ \citenamefont
  {Zubairy}(1997)}]{Scully}%
  \BibitemOpen
  \bibfield  {author} {\bibinfo {author} {\bibfnamefont {M.~O.}\ \bibnamefont
  {Scully}}\ and\ \bibinfo {author} {\bibfnamefont {M.~S.}\ \bibnamefont
  {Zubairy}},\ }\href@noop {} {\emph {\bibinfo {title} {Quantum Optics}}}\
  (\bibinfo  {publisher} {Cambridge University Press},\ \bibinfo {year}
  {1997})\BibitemShut {NoStop}%
\bibitem [{\citenamefont {Gardiner}\ and\ \citenamefont
  {Collett}(1985)}]{Gardiner1985}%
  \BibitemOpen
  \bibfield  {author} {\bibinfo {author} {\bibfnamefont {C.~W.}\ \bibnamefont
  {Gardiner}}\ and\ \bibinfo {author} {\bibfnamefont {M.~J.}\ \bibnamefont
  {Collett}},\ }\bibfield  {title} {\bibinfo {title} {Input and output in
  damped quantum systems: Quantum stochastic differential equations and the
  master equation},\ }\href {https://doi.org/10.1103/PhysRevA.31.3761}
  {\bibfield  {journal} {\bibinfo  {journal} {Phys. Rev. A}\ }\textbf {\bibinfo
  {volume} {31}},\ \bibinfo {pages} {3761} (\bibinfo {year}
  {1985})}\BibitemShut {NoStop}%
\bibitem [{\citenamefont {Zeytino\ifmmode~\check{g}\else \v{g}\fi{}lu}\ \emph
  {et~al.}(2017)\citenamefont {Zeytino\ifmmode~\check{g}\else \v{g}\fi{}lu},
  \citenamefont {Roth}, \citenamefont {Huber},\ and\ \citenamefont {\ifmmode
  \dot{I}\else \.{I}\fi{}mamo\ifmmode~\breve{g}\else
  \u{g}\fi{}lu}}]{Zeytinoglu2017}%
  \BibitemOpen
  \bibfield  {author} {\bibinfo {author} {\bibfnamefont {S.}~\bibnamefont
  {Zeytino\ifmmode~\check{g}\else \v{g}\fi{}lu}}, \bibinfo {author}
  {\bibfnamefont {C.}~\bibnamefont {Roth}}, \bibinfo {author} {\bibfnamefont
  {S.}~\bibnamefont {Huber}},\ and\ \bibinfo {author} {\bibfnamefont
  {A.}~\bibnamefont {\ifmmode \dot{I}\else
  \.{I}\fi{}mamo\ifmmode~\breve{g}\else \u{g}\fi{}lu}},\ }\bibfield  {title}
  {\bibinfo {title} {Atomically thin semiconductors as nonlinear mirrors},\
  }\href {https://doi.org/10.1103/PhysRevA.96.031801} {\bibfield  {journal}
  {\bibinfo  {journal} {Phys. Rev. A}\ }\textbf {\bibinfo {volume} {96}},\
  \bibinfo {pages} {031801} (\bibinfo {year} {2017})}\BibitemShut {NoStop}%
\bibitem [{\citenamefont {Meier}\ and\ \citenamefont
  {Zakharchenya}(1984)}]{OpticalOrientation}%
  \BibitemOpen
  \bibfield  {author} {\bibinfo {author} {\bibfnamefont {F.}~\bibnamefont
  {Meier}}\ and\ \bibinfo {author} {\bibfnamefont {B.}~\bibnamefont
  {Zakharchenya}},\ }\href@noop {} {\emph {\bibinfo {title} {Optical
  Orientation}}}\ (\bibinfo  {publisher} {North-Holland, Amsterdam},\ \bibinfo
  {year} {1984})\BibitemShut {NoStop}%
\bibitem [{\citenamefont {Pekar}(1958)}]{Pekar1958}%
  \BibitemOpen
  \bibfield  {author} {\bibinfo {author} {\bibfnamefont {S.~I.}\ \bibnamefont
  {Pekar}},\ }\bibfield  {title} {\bibinfo {title} {The energy of excitons for
  very small quasi-momenta},\ }\href@noop {} {\bibfield  {journal} {\bibinfo
  {journal} {Journal of Experimental and Theoretical Physics (U.S.S.R.)}\
  }\textbf {\bibinfo {volume} {35}},\ \bibinfo {pages} {522} (\bibinfo {year}
  {1958})}\BibitemShut {NoStop}%
\bibitem [{\citenamefont {Pikus}\ and\ \citenamefont {Bir}(1970)}]{Bir1970}%
  \BibitemOpen
  \bibfield  {author} {\bibinfo {author} {\bibfnamefont {G.~E.}\ \bibnamefont
  {Pikus}}\ and\ \bibinfo {author} {\bibfnamefont {G.~L.}\ \bibnamefont
  {Bir}},\ }\bibfield  {title} {\bibinfo {title} {Exchange interaction in
  excitons in semiconductors},\ }\href@noop {} {\bibfield  {journal} {\bibinfo
  {journal} {Sov. Phys. JETP}\ }\textbf {\bibinfo {volume} {33}},\ \bibinfo
  {pages} {195} (\bibinfo {year} {1970})}\BibitemShut {NoStop}%
\bibitem [{\citenamefont {Nakayama}(1985)}]{Nakayama1985}%
  \BibitemOpen
  \bibfield  {author} {\bibinfo {author} {\bibfnamefont {M.}~\bibnamefont
  {Nakayama}},\ }\bibfield  {title} {\bibinfo {title} {Theory of the excitonic
  polariton of the quantum well},\ }\href
  {https://doi.org/https://doi.org/10.1016/0038-1098(85)90131-0} {\bibfield
  {journal} {\bibinfo  {journal} {Solid State Communications}\ }\textbf
  {\bibinfo {volume} {55}},\ \bibinfo {pages} {1053} (\bibinfo {year}
  {1985})}\BibitemShut {NoStop}%
\bibitem [{\citenamefont {Cho}(1999)}]{Cho1999}%
  \BibitemOpen
  \bibfield  {author} {\bibinfo {author} {\bibfnamefont {K.}~\bibnamefont
  {Cho}},\ }\bibfield  {title} {\bibinfo {title} {Mechanisms for lt splitting
  of polarization waves: a link between electron-hole exchange interaction and
  depolarization shift},\ }\href {https://doi.org/10.1143/JPSJ.68.683}
  {\bibfield  {journal} {\bibinfo  {journal} {Journal of the Physical Society
  of Japan}\ }\textbf {\bibinfo {volume} {68}},\ \bibinfo {pages} {683}
  (\bibinfo {year} {1999})},\ \Eprint
  {https://arxiv.org/abs/https://doi.org/10.1143/JPSJ.68.683}
  {https://doi.org/10.1143/JPSJ.68.683} \BibitemShut {NoStop}%
\bibitem [{\citenamefont {Andreani}\ and\ \citenamefont
  {Bassani}(1990)}]{Andreani1990}%
  \BibitemOpen
  \bibfield  {author} {\bibinfo {author} {\bibfnamefont {L.~C.}\ \bibnamefont
  {Andreani}}\ and\ \bibinfo {author} {\bibfnamefont {F.}~\bibnamefont
  {Bassani}},\ }\bibfield  {title} {\bibinfo {title} {Exchange interaction and
  polariton effects in quantum-well excitons},\ }\href
  {https://doi.org/10.1103/PhysRevB.41.7536} {\bibfield  {journal} {\bibinfo
  {journal} {Phys. Rev. B}\ }\textbf {\bibinfo {volume} {41}},\ \bibinfo
  {pages} {7536} (\bibinfo {year} {1990})}\BibitemShut {NoStop}%
\bibitem [{\citenamefont {Maialle}\ \emph {et~al.}(1993)\citenamefont
  {Maialle}, \citenamefont {de~Andrada~e Silva},\ and\ \citenamefont
  {Sham}}]{Sham}%
  \BibitemOpen
  \bibfield  {author} {\bibinfo {author} {\bibfnamefont {M.~Z.}\ \bibnamefont
  {Maialle}}, \bibinfo {author} {\bibfnamefont {E.~A.}\ \bibnamefont
  {de~Andrada~e Silva}},\ and\ \bibinfo {author} {\bibfnamefont {L.~J.}\
  \bibnamefont {Sham}},\ }\bibfield  {title} {\bibinfo {title} {Exciton spin
  dynamics in quantum wells},\ }\href
  {https://doi.org/10.1103/PhysRevB.47.15776} {\bibfield  {journal} {\bibinfo
  {journal} {Phys. Rev. B}\ }\textbf {\bibinfo {volume} {47}},\ \bibinfo
  {pages} {15776} (\bibinfo {year} {1993})}\BibitemShut {NoStop}%
\bibitem [{\citenamefont {Gupalov}\ \emph {et~al.}(1998)\citenamefont
  {Gupalov}, \citenamefont {Ivchenko},\ and\ \citenamefont
  {Kavokin}}]{Gupalov}%
  \BibitemOpen
  \bibfield  {author} {\bibinfo {author} {\bibfnamefont {S.~V.}\ \bibnamefont
  {Gupalov}}, \bibinfo {author} {\bibfnamefont {E.~L.}\ \bibnamefont
  {Ivchenko}},\ and\ \bibinfo {author} {\bibfnamefont {A.~V.}\ \bibnamefont
  {Kavokin}},\ }\bibfield  {title} {\bibinfo {title} {Fine structure of
  localized exciton levels in quantum wells},\ }\href
  {https://doi.org/10.1134/1.558441} {\bibfield  {journal} {\bibinfo  {journal}
  {Journal of Experimental and Theoretical Physics}\ }\textbf {\bibinfo
  {volume} {86}},\ \bibinfo {pages} {388} (\bibinfo {year} {1998})}\BibitemShut
  {NoStop}%
\bibitem [{\citenamefont {Glazov}\ \emph {et~al.}(2014)\citenamefont {Glazov},
  \citenamefont {Amand}, \citenamefont {Marie}, \citenamefont {Lagarde},
  \citenamefont {Bouet},\ and\ \citenamefont {Urbaszek}}]{Glazov}%
  \BibitemOpen
  \bibfield  {author} {\bibinfo {author} {\bibfnamefont {M.~M.}\ \bibnamefont
  {Glazov}}, \bibinfo {author} {\bibfnamefont {T.}~\bibnamefont {Amand}},
  \bibinfo {author} {\bibfnamefont {X.}~\bibnamefont {Marie}}, \bibinfo
  {author} {\bibfnamefont {D.}~\bibnamefont {Lagarde}}, \bibinfo {author}
  {\bibfnamefont {L.}~\bibnamefont {Bouet}},\ and\ \bibinfo {author}
  {\bibfnamefont {B.}~\bibnamefont {Urbaszek}},\ }\bibfield  {title} {\bibinfo
  {title} {Exciton fine structure and spin decoherence in monolayers of
  transition metal dichalcogenides},\ }\href
  {https://doi.org/10.1103/PhysRevB.89.201302} {\bibfield  {journal} {\bibinfo
  {journal} {Phys. Rev. B}\ }\textbf {\bibinfo {volume} {89}},\ \bibinfo
  {pages} {201302(R)} (\bibinfo {year} {2014})}\BibitemShut {NoStop}%
\bibitem [{\citenamefont {Qiu}\ \emph {et~al.}(2015)\citenamefont {Qiu},
  \citenamefont {Cao},\ and\ \citenamefont {Louie}}]{Qiu2015}%
  \BibitemOpen
  \bibfield  {author} {\bibinfo {author} {\bibfnamefont {D.~Y.}\ \bibnamefont
  {Qiu}}, \bibinfo {author} {\bibfnamefont {T.}~\bibnamefont {Cao}},\ and\
  \bibinfo {author} {\bibfnamefont {S.~G.}\ \bibnamefont {Louie}},\ }\bibfield
  {title} {\bibinfo {title} {Nonanalyticity, valley quantum phases, and
  lightlike exciton dispersion in monolayer transition metal dichalcogenides:
  Theory and first-principles calculations},\ }\href
  {https://doi.org/10.1103/PhysRevLett.115.176801} {\bibfield  {journal}
  {\bibinfo  {journal} {Phys. Rev. Lett.}\ }\textbf {\bibinfo {volume} {115}},\
  \bibinfo {pages} {176801} (\bibinfo {year} {2015})}\BibitemShut {NoStop}%
\bibitem [{\citenamefont {Nestoklon}\ \emph {et~al.}(2018)\citenamefont
  {Nestoklon}, \citenamefont {Goupalov}, \citenamefont {Dzhioev}, \citenamefont
  {Ken}, \citenamefont {Korenev}, \citenamefont {Kusrayev}, \citenamefont
  {Sapega}, \citenamefont {de~Weerd}, \citenamefont {Gomez}, \citenamefont
  {Gregorkiewicz}, \citenamefont {Lin}, \citenamefont {Suenaga}, \citenamefont
  {Fujiwara}, \citenamefont {Matyushkin},\ and\ \citenamefont
  {Yassievich}}]{Nestoklon2018}%
  \BibitemOpen
  \bibfield  {author} {\bibinfo {author} {\bibfnamefont {M.~O.}\ \bibnamefont
  {Nestoklon}}, \bibinfo {author} {\bibfnamefont {S.~V.}\ \bibnamefont
  {Goupalov}}, \bibinfo {author} {\bibfnamefont {R.~I.}\ \bibnamefont
  {Dzhioev}}, \bibinfo {author} {\bibfnamefont {O.~S.}\ \bibnamefont {Ken}},
  \bibinfo {author} {\bibfnamefont {V.~L.}\ \bibnamefont {Korenev}}, \bibinfo
  {author} {\bibfnamefont {Y.~G.}\ \bibnamefont {Kusrayev}}, \bibinfo {author}
  {\bibfnamefont {V.~F.}\ \bibnamefont {Sapega}}, \bibinfo {author}
  {\bibfnamefont {C.}~\bibnamefont {de~Weerd}}, \bibinfo {author}
  {\bibfnamefont {L.}~\bibnamefont {Gomez}}, \bibinfo {author} {\bibfnamefont
  {T.}~\bibnamefont {Gregorkiewicz}}, \bibinfo {author} {\bibfnamefont
  {J.}~\bibnamefont {Lin}}, \bibinfo {author} {\bibfnamefont {K.}~\bibnamefont
  {Suenaga}}, \bibinfo {author} {\bibfnamefont {Y.}~\bibnamefont {Fujiwara}},
  \bibinfo {author} {\bibfnamefont {L.~B.}\ \bibnamefont {Matyushkin}},\ and\
  \bibinfo {author} {\bibfnamefont {I.~N.}\ \bibnamefont {Yassievich}},\
  }\bibfield  {title} {\bibinfo {title} {Optical orientation and alignment of
  excitons in ensembles of inorganic perovskite nanocrystals},\ }\href
  {https://doi.org/10.1103/PhysRevB.97.235304} {\bibfield  {journal} {\bibinfo
  {journal} {Phys. Rev. B}\ }\textbf {\bibinfo {volume} {97}},\ \bibinfo
  {pages} {235304} (\bibinfo {year} {2018})}\BibitemShut {NoStop}%
\bibitem [{\citenamefont {Becker}\ \emph {et~al.}(2018)\citenamefont {Becker},
  \citenamefont {Vaxenburg}, \citenamefont {Nedelcu}, \citenamefont {Sercel},
  \citenamefont {Shabaev}, \citenamefont {Mehl}, \citenamefont {Michopoulos},
  \citenamefont {Lambrakos}, \citenamefont {Bernstein}, \citenamefont {Lyons},
  \citenamefont {St{\"o}ferle}, \citenamefont {Mahrt}, \citenamefont
  {Kovalenko}, \citenamefont {Norris}, \citenamefont {Rain{\`o}},\ and\
  \citenamefont {Efros}}]{Becker2018}%
  \BibitemOpen
  \bibfield  {author} {\bibinfo {author} {\bibfnamefont {M.~A.}\ \bibnamefont
  {Becker}}, \bibinfo {author} {\bibfnamefont {R.}~\bibnamefont {Vaxenburg}},
  \bibinfo {author} {\bibfnamefont {G.}~\bibnamefont {Nedelcu}}, \bibinfo
  {author} {\bibfnamefont {P.~C.}\ \bibnamefont {Sercel}}, \bibinfo {author}
  {\bibfnamefont {A.}~\bibnamefont {Shabaev}}, \bibinfo {author} {\bibfnamefont
  {M.~J.}\ \bibnamefont {Mehl}}, \bibinfo {author} {\bibfnamefont {J.~G.}\
  \bibnamefont {Michopoulos}}, \bibinfo {author} {\bibfnamefont {S.~G.}\
  \bibnamefont {Lambrakos}}, \bibinfo {author} {\bibfnamefont {N.}~\bibnamefont
  {Bernstein}}, \bibinfo {author} {\bibfnamefont {J.~L.}\ \bibnamefont
  {Lyons}}, \bibinfo {author} {\bibfnamefont {T.}~\bibnamefont {St{\"o}ferle}},
  \bibinfo {author} {\bibfnamefont {R.~F.}\ \bibnamefont {Mahrt}}, \bibinfo
  {author} {\bibfnamefont {M.~V.}\ \bibnamefont {Kovalenko}}, \bibinfo {author}
  {\bibfnamefont {D.~J.}\ \bibnamefont {Norris}}, \bibinfo {author}
  {\bibfnamefont {G.}~\bibnamefont {Rain{\`o}}},\ and\ \bibinfo {author}
  {\bibfnamefont {A.~L.}\ \bibnamefont {Efros}},\ }\bibfield  {title} {\bibinfo
  {title} {Bright triplet excitons in caesium lead halide perovskites},\ }\href
  {https://doi.org/10.1038/nature25147} {\bibfield  {journal} {\bibinfo
  {journal} {Nature}\ }\textbf {\bibinfo {volume} {553}},\ \bibinfo {pages}
  {189} (\bibinfo {year} {2018})}\BibitemShut {NoStop}%
\bibitem [{\citenamefont {Andreev}(2021)}]{Andreev2021}%
  \BibitemOpen
  \bibfield  {author} {\bibinfo {author} {\bibfnamefont {S.~V.}\ \bibnamefont
  {Andreev}},\ }\bibfield  {title} {\bibinfo {title} {Spin-orbit-coupled
  depairing of a dipolar biexciton superfluid},\ }\href
  {https://doi.org/10.1103/PhysRevB.103.184503} {\bibfield  {journal} {\bibinfo
   {journal} {Phys. Rev. B}\ }\textbf {\bibinfo {volume} {103}},\ \bibinfo
  {pages} {184503} (\bibinfo {year} {2021})}\BibitemShut {NoStop}%
\bibitem [{\citenamefont {Andreev}(2022)}]{Andreev2022_1}%
  \BibitemOpen
  \bibfield  {author} {\bibinfo {author} {\bibfnamefont {S.~V.}\ \bibnamefont
  {Andreev}},\ }\bibfield  {title} {\bibinfo {title} {Pairing of
  electromagnetic bosons under spin-orbit coupling},\ }\href
  {https://doi.org/10.1103/PhysRevB.106.155157} {\bibfield  {journal} {\bibinfo
   {journal} {Phys. Rev. B}\ }\textbf {\bibinfo {volume} {106}},\ \bibinfo
  {pages} {155157} (\bibinfo {year} {2022})}\BibitemShut {NoStop}%
\bibitem [{\citenamefont {Klyshko}(1967)}]{Klyshko1967}%
  \BibitemOpen
  \bibfield  {author} {\bibinfo {author} {\bibfnamefont {D.~N.}\ \bibnamefont
  {Klyshko}},\ }\bibfield  {title} {\bibinfo {title} {Coherent photon decay in
  a nonlinear medium},\ }\href@noop {} {\bibfield  {journal} {\bibinfo
  {journal} {JETP Letters}\ }\textbf {\bibinfo {volume} {6}},\ \bibinfo {pages}
  {23} (\bibinfo {year} {1967})}\BibitemShut {NoStop}%
\bibitem [{\citenamefont {Heidmann}\ \emph {et~al.}(1987)\citenamefont
  {Heidmann}, \citenamefont {Horowicz}, \citenamefont {Reynaud}, \citenamefont
  {Giacobino}, \citenamefont {Fabre},\ and\ \citenamefont
  {Camy}}]{Heidmann1987}%
  \BibitemOpen
  \bibfield  {author} {\bibinfo {author} {\bibfnamefont {A.}~\bibnamefont
  {Heidmann}}, \bibinfo {author} {\bibfnamefont {R.~J.}\ \bibnamefont
  {Horowicz}}, \bibinfo {author} {\bibfnamefont {S.}~\bibnamefont {Reynaud}},
  \bibinfo {author} {\bibfnamefont {E.}~\bibnamefont {Giacobino}}, \bibinfo
  {author} {\bibfnamefont {C.}~\bibnamefont {Fabre}},\ and\ \bibinfo {author}
  {\bibfnamefont {G.}~\bibnamefont {Camy}},\ }\bibfield  {title} {\bibinfo
  {title} {Observation of quantum noise reduction on twin laser beams},\ }\href
  {https://doi.org/10.1103/PhysRevLett.59.2555} {\bibfield  {journal} {\bibinfo
   {journal} {Phys. Rev. Lett.}\ }\textbf {\bibinfo {volume} {59}},\ \bibinfo
  {pages} {2555} (\bibinfo {year} {1987})}\BibitemShut {NoStop}%
\bibitem [{\citenamefont {Zel'dovich}\ \emph {et~al.}(1985)\citenamefont
  {Zel'dovich}, \citenamefont {Pilipetsky},\ and\ \citenamefont
  {Shkunov}}]{Zeldovich1985}%
  \BibitemOpen
  \bibfield  {author} {\bibinfo {author} {\bibfnamefont {B.~Y.}\ \bibnamefont
  {Zel'dovich}}, \bibinfo {author} {\bibfnamefont {N.~F.}\ \bibnamefont
  {Pilipetsky}},\ and\ \bibinfo {author} {\bibfnamefont {V.~V.}\ \bibnamefont
  {Shkunov}},\ }\href@noop {} {\emph {\bibinfo {title} {Principles of Phase
  Conjugation}}},\ \bibinfo {series} {Springer Series in Optical Sciences},
  Vol.~\bibinfo {volume} {42}\ (\bibinfo  {publisher} {Springer Berlin,
  Heidelberg},\ \bibinfo {year} {1985})\BibitemShut {NoStop}%
\bibitem [{\citenamefont {Elliott}(1957)}]{Elliott1957}%
  \BibitemOpen
  \bibfield  {author} {\bibinfo {author} {\bibfnamefont {R.~J.}\ \bibnamefont
  {Elliott}},\ }\bibfield  {title} {\bibinfo {title} {Intensity of optical
  absorption by excitons},\ }\href {https://doi.org/10.1103/PhysRev.108.1384}
  {\bibfield  {journal} {\bibinfo  {journal} {Phys. Rev.}\ }\textbf {\bibinfo
  {volume} {108}},\ \bibinfo {pages} {1384} (\bibinfo {year}
  {1957})}\BibitemShut {NoStop}%
\bibitem [{\citenamefont {Xiao}\ \emph {et~al.}(2012)\citenamefont {Xiao},
  \citenamefont {Liu}, \citenamefont {Feng}, \citenamefont {Xu},\ and\
  \citenamefont {Yao}}]{Xiao2012}%
  \BibitemOpen
  \bibfield  {author} {\bibinfo {author} {\bibfnamefont {D.}~\bibnamefont
  {Xiao}}, \bibinfo {author} {\bibfnamefont {G.-B.}\ \bibnamefont {Liu}},
  \bibinfo {author} {\bibfnamefont {W.}~\bibnamefont {Feng}}, \bibinfo {author}
  {\bibfnamefont {X.}~\bibnamefont {Xu}},\ and\ \bibinfo {author}
  {\bibfnamefont {W.}~\bibnamefont {Yao}},\ }\bibfield  {title} {\bibinfo
  {title} {Coupled spin and valley physics in monolayers of
  ${\mathrm{mos}}_{2}$ and other group-vi dichalcogenides},\ }\href
  {https://doi.org/10.1103/PhysRevLett.108.196802} {\bibfield  {journal}
  {\bibinfo  {journal} {Phys. Rev. Lett.}\ }\textbf {\bibinfo {volume} {108}},\
  \bibinfo {pages} {196802} (\bibinfo {year} {2012})}\BibitemShut {NoStop}%
\bibitem [{\citenamefont {Zeng}\ \emph {et~al.}(2012)\citenamefont {Zeng},
  \citenamefont {Dai}, \citenamefont {Yao}, \citenamefont {Xiao},\ and\
  \citenamefont {Cui}}]{Zeng2012}%
  \BibitemOpen
  \bibfield  {author} {\bibinfo {author} {\bibfnamefont {H.}~\bibnamefont
  {Zeng}}, \bibinfo {author} {\bibfnamefont {J.}~\bibnamefont {Dai}}, \bibinfo
  {author} {\bibfnamefont {W.}~\bibnamefont {Yao}}, \bibinfo {author}
  {\bibfnamefont {D.}~\bibnamefont {Xiao}},\ and\ \bibinfo {author}
  {\bibfnamefont {X.}~\bibnamefont {Cui}},\ }\bibfield  {title} {\bibinfo
  {title} {Valley polarization in mos2 monolayers by optical pumping},\ }\href
  {https://doi.org/10.1038/nnano.2012.95} {\bibfield  {journal} {\bibinfo
  {journal} {Nature Nanotechnology}\ }\textbf {\bibinfo {volume} {7}},\
  \bibinfo {pages} {490} (\bibinfo {year} {2012})}\BibitemShut {NoStop}%
\bibitem [{\citenamefont {Sallen}\ \emph {et~al.}(2012)\citenamefont {Sallen},
  \citenamefont {Bouet}, \citenamefont {Marie}, \citenamefont {Wang},
  \citenamefont {Zhu}, \citenamefont {Han}, \citenamefont {Lu}, \citenamefont
  {Tan}, \citenamefont {Amand}, \citenamefont {Liu},\ and\ \citenamefont
  {Urbaszek}}]{Sallen2012}%
  \BibitemOpen
  \bibfield  {author} {\bibinfo {author} {\bibfnamefont {G.}~\bibnamefont
  {Sallen}}, \bibinfo {author} {\bibfnamefont {L.}~\bibnamefont {Bouet}},
  \bibinfo {author} {\bibfnamefont {X.}~\bibnamefont {Marie}}, \bibinfo
  {author} {\bibfnamefont {G.}~\bibnamefont {Wang}}, \bibinfo {author}
  {\bibfnamefont {C.~R.}\ \bibnamefont {Zhu}}, \bibinfo {author} {\bibfnamefont
  {W.~P.}\ \bibnamefont {Han}}, \bibinfo {author} {\bibfnamefont
  {Y.}~\bibnamefont {Lu}}, \bibinfo {author} {\bibfnamefont {P.~H.}\
  \bibnamefont {Tan}}, \bibinfo {author} {\bibfnamefont {T.}~\bibnamefont
  {Amand}}, \bibinfo {author} {\bibfnamefont {B.~L.}\ \bibnamefont {Liu}},\
  and\ \bibinfo {author} {\bibfnamefont {B.}~\bibnamefont {Urbaszek}},\
  }\bibfield  {title} {\bibinfo {title} {Robust optical emission polarization
  in mos${}_{2}$ monolayers through selective valley excitation},\ }\href
  {https://doi.org/10.1103/PhysRevB.86.081301} {\bibfield  {journal} {\bibinfo
  {journal} {Phys. Rev. B}\ }\textbf {\bibinfo {volume} {86}},\ \bibinfo
  {pages} {081301} (\bibinfo {year} {2012})}\BibitemShut {NoStop}%
\bibitem [{\citenamefont {Srivastava}\ \emph {et~al.}(2015)\citenamefont
  {Srivastava}, \citenamefont {Sidler}, \citenamefont {Allain}, \citenamefont
  {Lembke}, \citenamefont {Kis},\ and\ \citenamefont {Imamo{\u
  g}lu}}]{Srivastava2015}%
  \BibitemOpen
  \bibfield  {author} {\bibinfo {author} {\bibfnamefont {A.}~\bibnamefont
  {Srivastava}}, \bibinfo {author} {\bibfnamefont {M.}~\bibnamefont {Sidler}},
  \bibinfo {author} {\bibfnamefont {A.~V.}\ \bibnamefont {Allain}}, \bibinfo
  {author} {\bibfnamefont {D.~S.}\ \bibnamefont {Lembke}}, \bibinfo {author}
  {\bibfnamefont {A.}~\bibnamefont {Kis}},\ and\ \bibinfo {author}
  {\bibfnamefont {A.}~\bibnamefont {Imamo{\u g}lu}},\ }\bibfield  {title}
  {\bibinfo {title} {Valley zeeman effect in elementary optical excitations of
  monolayer wse2},\ }\href {https://doi.org/10.1038/nphys3203} {\bibfield
  {journal} {\bibinfo  {journal} {Nature Physics}\ }\textbf {\bibinfo {volume}
  {11}},\ \bibinfo {pages} {141} (\bibinfo {year} {2015})}\BibitemShut
  {NoStop}%
\bibitem [{\citenamefont {Stier}\ \emph {et~al.}(2016)\citenamefont {Stier},
  \citenamefont {McCreary}, \citenamefont {Jonker}, \citenamefont {Kono},\ and\
  \citenamefont {Crooker}}]{Stier2016}%
  \BibitemOpen
  \bibfield  {author} {\bibinfo {author} {\bibfnamefont {A.~V.}\ \bibnamefont
  {Stier}}, \bibinfo {author} {\bibfnamefont {K.~M.}\ \bibnamefont {McCreary}},
  \bibinfo {author} {\bibfnamefont {B.~T.}\ \bibnamefont {Jonker}}, \bibinfo
  {author} {\bibfnamefont {J.}~\bibnamefont {Kono}},\ and\ \bibinfo {author}
  {\bibfnamefont {S.~A.}\ \bibnamefont {Crooker}},\ }\bibfield  {title}
  {\bibinfo {title} {Exciton diamagnetic shifts and valley zeeman effects in
  monolayer ws2 and mos2 to 65 tesla},\ }\href
  {https://doi.org/10.1038/ncomms10643} {\bibfield  {journal} {\bibinfo
  {journal} {Nature Communications}\ }\textbf {\bibinfo {volume} {7}},\
  \bibinfo {pages} {10643} (\bibinfo {year} {2016})}\BibitemShut {NoStop}%
\bibitem [{\citenamefont {Wang}\ \emph {et~al.}(2026)\citenamefont {Wang},
  \citenamefont {Kim}, \citenamefont {Zhen},\ and\ \citenamefont
  {He}}]{Wang2026}%
  \BibitemOpen
  \bibfield  {author} {\bibinfo {author} {\bibfnamefont {Z.}~\bibnamefont
  {Wang}}, \bibinfo {author} {\bibfnamefont {B.}~\bibnamefont {Kim}}, \bibinfo
  {author} {\bibfnamefont {B.}~\bibnamefont {Zhen}},\ and\ \bibinfo {author}
  {\bibfnamefont {L.}~\bibnamefont {He}},\ }\bibfield  {title} {\bibinfo
  {title} {Strongly nonlinear nanocavity exciton polaritons in gate-tunable
  monolayer semiconductors},\ }\href {https://doi.org/10.1103/gc15-qsvf}
  {\bibfield  {journal} {\bibinfo  {journal} {Phys. Rev. Lett.}\ }\textbf
  {\bibinfo {volume} {136}},\ \bibinfo {pages} {146901} (\bibinfo {year}
  {2026})}\BibitemShut {NoStop}%
\bibitem [{\citenamefont {Utesov}\ and\ \citenamefont
  {Andreev}(2024)}]{Utesov2024}%
  \BibitemOpen
  \bibfield  {author} {\bibinfo {author} {\bibfnamefont {O.~I.}\ \bibnamefont
  {Utesov}}\ and\ \bibinfo {author} {\bibfnamefont {S.~V.}\ \bibnamefont
  {Andreev}},\ }\bibfield  {title} {\bibinfo {title} {Magnetoroton in a
  two-dimensional bose-bose mixture},\ }\href
  {https://doi.org/10.1103/PhysRevB.109.235135} {\bibfield  {journal} {\bibinfo
   {journal} {Phys. Rev. B}\ }\textbf {\bibinfo {volume} {109}},\ \bibinfo
  {pages} {235135} (\bibinfo {year} {2024})}\BibitemShut {NoStop}%
\bibitem [{\citenamefont {Andreev}(2024)}]{Andreev2024}%
  \BibitemOpen
  \bibfield  {author} {\bibinfo {author} {\bibfnamefont {S.~V.}\ \bibnamefont
  {Andreev}},\ }\bibfield  {title} {\bibinfo {title} {Controllable fusion of
  electromagnetic bosons in two-dimensional semiconductors},\ }\href
  {https://doi.org/10.1103/PhysRevB.109.205423} {\bibfield  {journal} {\bibinfo
   {journal} {Phys. Rev. B}\ }\textbf {\bibinfo {volume} {109}},\ \bibinfo
  {pages} {205423} (\bibinfo {year} {2024})}\BibitemShut {NoStop}%
\bibitem [{\citenamefont {Hanamura}(1988)}]{Hanamura1988}%
  \BibitemOpen
  \bibfield  {author} {\bibinfo {author} {\bibfnamefont {E.}~\bibnamefont
  {Hanamura}},\ }\bibfield  {title} {\bibinfo {title} {Rapid radiative decay
  and enhanced optical nonlinearity of excitons in a quantum well},\ }\href
  {https://doi.org/10.1103/PhysRevB.38.1228} {\bibfield  {journal} {\bibinfo
  {journal} {Phys. Rev. B}\ }\textbf {\bibinfo {volume} {38}},\ \bibinfo
  {pages} {1228} (\bibinfo {year} {1988})}\BibitemShut {NoStop}%
\bibitem [{\citenamefont {Liu}\ \emph {et~al.}(2026)\citenamefont {Liu},
  \citenamefont {Woo}, \citenamefont {Wu}, \citenamefont {Hou}, \citenamefont
  {Su},\ and\ \citenamefont {Qiu}}]{Liu2026}%
  \BibitemOpen
  \bibfield  {author} {\bibinfo {author} {\bibfnamefont {L.~Y.}\ \bibnamefont
  {Liu}}, \bibinfo {author} {\bibfnamefont {S.~Y.}\ \bibnamefont {Woo}},
  \bibinfo {author} {\bibfnamefont {J.}~\bibnamefont {Wu}}, \bibinfo {author}
  {\bibfnamefont {B.}~\bibnamefont {Hou}}, \bibinfo {author} {\bibfnamefont
  {C.}~\bibnamefont {Su}},\ and\ \bibinfo {author} {\bibfnamefont {D.~Y.}\
  \bibnamefont {Qiu}},\ }\bibfield  {title} {\bibinfo {title} {Direct
  observation of massless excitons and linear exciton dispersion},\ }\href
  {https://doi.org/10.1038/s41567-026-03193-8} {\bibfield  {journal} {\bibinfo
  {journal} {Nature Physics}\ }\textbf {\bibinfo {volume} {22}},\ \bibinfo
  {pages} {521} (\bibinfo {year} {2026})}\BibitemShut {NoStop}%
\bibitem [{\citenamefont {Tsesses}\ \emph {et~al.}(2018)\citenamefont
  {Tsesses}, \citenamefont {Ostrovsky}, \citenamefont {Cohen}, \citenamefont
  {Gjonaj}, \citenamefont {Lindner},\ and\ \citenamefont
  {Bartal}}]{Tsesses2018}%
  \BibitemOpen
  \bibfield  {author} {\bibinfo {author} {\bibfnamefont {S.}~\bibnamefont
  {Tsesses}}, \bibinfo {author} {\bibfnamefont {E.}~\bibnamefont {Ostrovsky}},
  \bibinfo {author} {\bibfnamefont {K.}~\bibnamefont {Cohen}}, \bibinfo
  {author} {\bibfnamefont {B.}~\bibnamefont {Gjonaj}}, \bibinfo {author}
  {\bibfnamefont {N.~H.}\ \bibnamefont {Lindner}},\ and\ \bibinfo {author}
  {\bibfnamefont {G.}~\bibnamefont {Bartal}},\ }\bibfield  {title} {\bibinfo
  {title} {Optical skyrmion lattice in evanescent electromagnetic fields},\
  }\href {https://www.science.org/doi/abs/10.1126/science.aau0227} {\bibfield
  {journal} {\bibinfo  {journal} {Science}\ }\textbf {\bibinfo {volume}
  {361}},\ \bibinfo {pages} {993} (\bibinfo {year} {2018})}\BibitemShut
  {NoStop}%
\bibitem [{\citenamefont {Hao}\ \emph {et~al.}(2025)\citenamefont {Hao},
  \citenamefont {Owrutsky}, \citenamefont {Ratchford}, \citenamefont
  {Simpkins},\ and\ \citenamefont {Efros}}]{Hao2025}%
  \BibitemOpen
  \bibfield  {author} {\bibinfo {author} {\bibfnamefont {J.}~\bibnamefont
  {Hao}}, \bibinfo {author} {\bibfnamefont {J.}~\bibnamefont {Owrutsky}},
  \bibinfo {author} {\bibfnamefont {D.~C.}\ \bibnamefont {Ratchford}}, \bibinfo
  {author} {\bibfnamefont {B.}~\bibnamefont {Simpkins}},\ and\ \bibinfo
  {author} {\bibfnamefont {A.~L.}\ \bibnamefont {Efros}},\ }\bibfield  {title}
  {\bibinfo {title} {Surface exciton polariton in cesium lead halide
  perovskites},\ }\href {https://doi.org/10.1021/acs.chemmater.4c03262}
  {\bibfield  {journal} {\bibinfo  {journal} {Chemistry of Materials}\ }\textbf
  {\bibinfo {volume} {37}},\ \bibinfo {pages} {2251} (\bibinfo {year}
  {2025})},\ \Eprint
  {https://arxiv.org/abs/https://pubs.acs.org/cmatex/article-pdf/37/6/2251/41184647/cm4c03262.pdf}
  {https://pubs.acs.org/cmatex/article-pdf/37/6/2251/41184647/cm4c03262.pdf}
  \BibitemShut {NoStop}%
\bibitem [{\citenamefont {Zhou}\ \emph {et~al.}(2017)\citenamefont {Zhou},
  \citenamefont {Scuri}, \citenamefont {Wild}, \citenamefont {High},
  \citenamefont {Dibos}, \citenamefont {Jauregui}, \citenamefont {Shu},
  \citenamefont {De~Greve}, \citenamefont {Pistunova}, \citenamefont {Joe},
  \citenamefont {Taniguchi}, \citenamefont {Watanabe}, \citenamefont {Kim},
  \citenamefont {Lukin},\ and\ \citenamefont {Park}}]{Zhou2017}%
  \BibitemOpen
  \bibfield  {author} {\bibinfo {author} {\bibfnamefont {Y.}~\bibnamefont
  {Zhou}}, \bibinfo {author} {\bibfnamefont {G.}~\bibnamefont {Scuri}},
  \bibinfo {author} {\bibfnamefont {D.~S.}\ \bibnamefont {Wild}}, \bibinfo
  {author} {\bibfnamefont {A.~A.}\ \bibnamefont {High}}, \bibinfo {author}
  {\bibfnamefont {A.}~\bibnamefont {Dibos}}, \bibinfo {author} {\bibfnamefont
  {L.~A.}\ \bibnamefont {Jauregui}}, \bibinfo {author} {\bibfnamefont
  {C.}~\bibnamefont {Shu}}, \bibinfo {author} {\bibfnamefont {K.}~\bibnamefont
  {De~Greve}}, \bibinfo {author} {\bibfnamefont {K.}~\bibnamefont {Pistunova}},
  \bibinfo {author} {\bibfnamefont {A.~Y.}\ \bibnamefont {Joe}}, \bibinfo
  {author} {\bibfnamefont {T.}~\bibnamefont {Taniguchi}}, \bibinfo {author}
  {\bibfnamefont {K.}~\bibnamefont {Watanabe}}, \bibinfo {author}
  {\bibfnamefont {P.}~\bibnamefont {Kim}}, \bibinfo {author} {\bibfnamefont
  {M.~D.}\ \bibnamefont {Lukin}},\ and\ \bibinfo {author} {\bibfnamefont
  {H.}~\bibnamefont {Park}},\ }\bibfield  {title} {\bibinfo {title} {Probing
  dark excitons in atomically thin semiconductors via near-field coupling to
  surface plasmon polaritons},\ }\href {https://doi.org/10.1038/nnano.2017.106}
  {\bibfield  {journal} {\bibinfo  {journal} {Nature Nanotechnology}\ }\textbf
  {\bibinfo {volume} {12}},\ \bibinfo {pages} {856} (\bibinfo {year}
  {2017})}\BibitemShut {NoStop}%
\bibitem [{\citenamefont {Jin}\ \emph {et~al.}(2026)\citenamefont {Jin},
  \citenamefont {Liu}, \citenamefont {Razdolski}, \citenamefont {Lo},
  \citenamefont {Wang}, \citenamefont {Peng}, \citenamefont {Liang},
  \citenamefont {Zhu}, \citenamefont {Yao}, \citenamefont {Zayats},\ and\
  \citenamefont {Lei}}]{Jin2026}%
  \BibitemOpen
  \bibfield  {author} {\bibinfo {author} {\bibfnamefont {S.}~\bibnamefont
  {Jin}}, \bibinfo {author} {\bibfnamefont {F.}~\bibnamefont {Liu}}, \bibinfo
  {author} {\bibfnamefont {I.}~\bibnamefont {Razdolski}}, \bibinfo {author}
  {\bibfnamefont {T.~W.}\ \bibnamefont {Lo}}, \bibinfo {author} {\bibfnamefont
  {Y.}~\bibnamefont {Wang}}, \bibinfo {author} {\bibfnamefont {Z.}~\bibnamefont
  {Peng}}, \bibinfo {author} {\bibfnamefont {K.}~\bibnamefont {Liang}},
  \bibinfo {author} {\bibfnamefont {Y.}~\bibnamefont {Zhu}}, \bibinfo {author}
  {\bibfnamefont {W.}~\bibnamefont {Yao}}, \bibinfo {author} {\bibfnamefont
  {A.~V.}\ \bibnamefont {Zayats}},\ and\ \bibinfo {author} {\bibfnamefont
  {D.}~\bibnamefont {Lei}},\ }\bibfield  {title} {\bibinfo {title} {Plasmonic
  tuning of dark-exciton radiation dynamics and far-field emission
  directionality in monolayer wse<sub>2</sub>},\ }\href
  {https://www.science.org/doi/abs/10.1126/sciadv.aea5781} {\bibfield
  {journal} {\bibinfo  {journal} {Science Advances}\ }\textbf {\bibinfo
  {volume} {12}},\ \bibinfo {pages} {eaea5781} (\bibinfo {year}
  {2026})}\BibitemShut {NoStop}%
\bibitem [{\citenamefont {Chen}\ \emph {et~al.}(2013)\citenamefont {Chen},
  \citenamefont {Nesterov}, \citenamefont {Nikitin}, \citenamefont
  {Thongrattanasiri}, \citenamefont {Alonso-Gonz{\'a}lez}, \citenamefont
  {Slipchenko}, \citenamefont {Speck}, \citenamefont {Ostler}, \citenamefont
  {Seyller}, \citenamefont {Crassee}, \citenamefont {Koppens}, \citenamefont
  {Martin-Moreno}, \citenamefont {Garc{\'\i}a~de Abajo}, \citenamefont
  {Kuzmenko},\ and\ \citenamefont {Hillenbrand}}]{Nikitin2013}%
  \BibitemOpen
  \bibfield  {author} {\bibinfo {author} {\bibfnamefont {J.}~\bibnamefont
  {Chen}}, \bibinfo {author} {\bibfnamefont {M.~L.}\ \bibnamefont {Nesterov}},
  \bibinfo {author} {\bibfnamefont {A.~Y.}\ \bibnamefont {Nikitin}}, \bibinfo
  {author} {\bibfnamefont {S.}~\bibnamefont {Thongrattanasiri}}, \bibinfo
  {author} {\bibfnamefont {P.}~\bibnamefont {Alonso-Gonz{\'a}lez}}, \bibinfo
  {author} {\bibfnamefont {T.~M.}\ \bibnamefont {Slipchenko}}, \bibinfo
  {author} {\bibfnamefont {F.}~\bibnamefont {Speck}}, \bibinfo {author}
  {\bibfnamefont {M.}~\bibnamefont {Ostler}}, \bibinfo {author} {\bibfnamefont
  {T.}~\bibnamefont {Seyller}}, \bibinfo {author} {\bibfnamefont
  {I.}~\bibnamefont {Crassee}}, \bibinfo {author} {\bibfnamefont {F.~H.~L.}\
  \bibnamefont {Koppens}}, \bibinfo {author} {\bibfnamefont {L.}~\bibnamefont
  {Martin-Moreno}}, \bibinfo {author} {\bibfnamefont {F.~J.}\ \bibnamefont
  {Garc{\'\i}a~de Abajo}}, \bibinfo {author} {\bibfnamefont {A.~B.}\
  \bibnamefont {Kuzmenko}},\ and\ \bibinfo {author} {\bibfnamefont
  {R.}~\bibnamefont {Hillenbrand}},\ }\bibfield  {title} {\bibinfo {title}
  {Strong plasmon reflection at nanometer-size gaps in monolayer graphene on
  sic},\ }\href {https://doi.org/10.1021/nl403622t} {\bibfield  {journal}
  {\bibinfo  {journal} {Nano Letters}\ }\textbf {\bibinfo {volume} {13}},\
  \bibinfo {pages} {6210} (\bibinfo {year} {2013})},\ \Eprint
  {https://arxiv.org/abs/https://pubs.acs.org/nalefd/article-pdf/13/12/6210/18346549/nl403622t.pdf}
  {https://pubs.acs.org/nalefd/article-pdf/13/12/6210/18346549/nl403622t.pdf}
  \BibitemShut {NoStop}%
\bibitem [{\citenamefont {S\'anchez-Gil}\ and\ \citenamefont
  {Maradudin}(1999)}]{Maradudin1999}%
  \BibitemOpen
  \bibfield  {author} {\bibinfo {author} {\bibfnamefont {J.~A.}\ \bibnamefont
  {S\'anchez-Gil}}\ and\ \bibinfo {author} {\bibfnamefont {A.~A.}\ \bibnamefont
  {Maradudin}},\ }\bibfield  {title} {\bibinfo {title} {Near-field and
  far-field scattering of surface plasmon polaritons by one-dimensional surface
  defects},\ }\href {https://doi.org/10.1103/PhysRevB.60.8359} {\bibfield
  {journal} {\bibinfo  {journal} {Phys. Rev. B}\ }\textbf {\bibinfo {volume}
  {60}},\ \bibinfo {pages} {8359} (\bibinfo {year} {1999})}\BibitemShut
  {NoStop}%
\bibitem [{\citenamefont {Nikitin}\ \emph {et~al.}(2007)\citenamefont
  {Nikitin}, \citenamefont {L\'opez-Tejeira},\ and\ \citenamefont
  {Mart\'{\i}n-Moreno}}]{Nikitin2007}%
  \BibitemOpen
  \bibfield  {author} {\bibinfo {author} {\bibfnamefont {A.~Y.}\ \bibnamefont
  {Nikitin}}, \bibinfo {author} {\bibfnamefont {F.}~\bibnamefont
  {L\'opez-Tejeira}},\ and\ \bibinfo {author} {\bibfnamefont {L.}~\bibnamefont
  {Mart\'{\i}n-Moreno}},\ }\bibfield  {title} {\bibinfo {title} {Scattering of
  surface plasmon polaritons by one-dimensional inhomogeneities},\ }\href
  {https://doi.org/10.1103/PhysRevB.75.035129} {\bibfield  {journal} {\bibinfo
  {journal} {Phys. Rev. B}\ }\textbf {\bibinfo {volume} {75}},\ \bibinfo
  {pages} {035129} (\bibinfo {year} {2007})}\BibitemShut {NoStop}%
\bibitem [{\citenamefont {Rejaei}\ and\ \citenamefont
  {Khavasi}(2015)}]{Rejaei2015}%
  \BibitemOpen
  \bibfield  {author} {\bibinfo {author} {\bibfnamefont {B.}~\bibnamefont
  {Rejaei}}\ and\ \bibinfo {author} {\bibfnamefont {A.}~\bibnamefont
  {Khavasi}},\ }\bibfield  {title} {\bibinfo {title} {Scattering of surface
  plasmons on graphene by a discontinuity in surface conductivity},\ }\href
  {https://doi.org/10.1088/2040-8978/17/7/075002} {\bibfield  {journal}
  {\bibinfo  {journal} {Journal of Optics}\ }\textbf {\bibinfo {volume} {17}},\
  \bibinfo {pages} {075002} (\bibinfo {year} {2015})}\BibitemShut {NoStop}%
\bibitem [{\citenamefont {del {\'A}guila}\ \emph {et~al.}(2023)\citenamefont
  {del {\'A}guila}, \citenamefont {Wong}, \citenamefont {Wadgaonkar},
  \citenamefont {Fieramosca}, \citenamefont {Liu}, \citenamefont {Vaklinova},
  \citenamefont {Dal~Forno}, \citenamefont {Do}, \citenamefont {Wei},
  \citenamefont {Watanabe}, \citenamefont {Taniguchi}, \citenamefont
  {Novoselov}, \citenamefont {Koperski}, \citenamefont {Battiato},\ and\
  \citenamefont {Xiong}}]{Aguila2023}%
  \BibitemOpen
  \bibfield  {author} {\bibinfo {author} {\bibfnamefont {A.~G.}\ \bibnamefont
  {del {\'A}guila}}, \bibinfo {author} {\bibfnamefont {Y.~R.}\ \bibnamefont
  {Wong}}, \bibinfo {author} {\bibfnamefont {I.}~\bibnamefont {Wadgaonkar}},
  \bibinfo {author} {\bibfnamefont {A.}~\bibnamefont {Fieramosca}}, \bibinfo
  {author} {\bibfnamefont {X.}~\bibnamefont {Liu}}, \bibinfo {author}
  {\bibfnamefont {K.}~\bibnamefont {Vaklinova}}, \bibinfo {author}
  {\bibfnamefont {S.}~\bibnamefont {Dal~Forno}}, \bibinfo {author}
  {\bibfnamefont {T.~T.~H.}\ \bibnamefont {Do}}, \bibinfo {author}
  {\bibfnamefont {H.~Y.}\ \bibnamefont {Wei}}, \bibinfo {author} {\bibfnamefont
  {K.}~\bibnamefont {Watanabe}}, \bibinfo {author} {\bibfnamefont
  {T.}~\bibnamefont {Taniguchi}}, \bibinfo {author} {\bibfnamefont {K.~S.}\
  \bibnamefont {Novoselov}}, \bibinfo {author} {\bibfnamefont {M.}~\bibnamefont
  {Koperski}}, \bibinfo {author} {\bibfnamefont {M.}~\bibnamefont {Battiato}},\
  and\ \bibinfo {author} {\bibfnamefont {Q.}~\bibnamefont {Xiong}},\ }\bibfield
   {title} {\bibinfo {title} {Ultrafast exciton fluid flow in an atomically
  thin mos2 semiconductor},\ }\href
  {https://doi.org/10.1038/s41565-023-01438-8} {\bibfield  {journal} {\bibinfo
  {journal} {Nature Nanotechnology}\ }\textbf {\bibinfo {volume} {18}},\
  \bibinfo {pages} {1012} (\bibinfo {year} {2023})}\BibitemShut {NoStop}%
\bibitem [{\citenamefont {Harris}(1966)}]{Harris1966}%
  \BibitemOpen
  \bibfield  {author} {\bibinfo {author} {\bibfnamefont {S.~E.}\ \bibnamefont
  {Harris}},\ }\bibfield  {title} {\bibinfo {title} {Proposed backward wave
  oscillation in the infrared},\ }\bibfield  {booktitle} {\emph {\bibinfo
  {booktitle} {Applied Physics Letters}},\ }\href
  {https://doi.org/10.1063/1.1754668} {\bibfield  {journal} {\bibinfo
  {journal} {Applied Physics Letters}\ }\textbf {\bibinfo {volume} {9}},\
  \bibinfo {pages} {114} (\bibinfo {year} {1966})}\BibitemShut {NoStop}%
\bibitem [{\citenamefont {Canalias}\ and\ \citenamefont
  {Pasiskevicius}(2007)}]{Canalias2007}%
  \BibitemOpen
  \bibfield  {author} {\bibinfo {author} {\bibfnamefont {C.}~\bibnamefont
  {Canalias}}\ and\ \bibinfo {author} {\bibfnamefont {V.}~\bibnamefont
  {Pasiskevicius}},\ }\bibfield  {title} {\bibinfo {title} {Mirrorless optical
  parametric oscillator},\ }\href {https://doi.org/10.1038/nphoton.2007.137}
  {\bibfield  {journal} {\bibinfo  {journal} {Nature Photonics}\ }\textbf
  {\bibinfo {volume} {1}},\ \bibinfo {pages} {459} (\bibinfo {year}
  {2007})}\BibitemShut {NoStop}%
\end{thebibliography}%
\end{document}